\documentclass[a4paper,11pt]{book}
\usepackage{graphicx}
\usepackage[USenglish,shorthands=off]{babel}
\usepackage{amsmath,amssymb,amsthm}
\usepackage{ulem}
\usepackage{bm}
\usepackage[authoryear]{natbib}
\usepackage{indentfirst}
\usepackage[labelfont=bf]{caption}
\usepackage{wrapfig}
\usepackage{hyperref}
\makeatletter
\newcommand{\leqnomode}{\tagsleft@true}
\newcommand{\reqnomode}{\tagsleft@false}
\makeatother
\DeclareSymbolFont{symbolsC}{U}{pxsyc}{m}{n}
\DeclareMathSymbol{\coloneqq}{\mathrel}{symbolsC}{"42}
\theoremstyle{definition}

\makeatletter
\def\blfootnote{\xdef\@thefnmark{}\@footnotetext}
\makeatother
\renewcommand{\bibname}{References}

\renewcommand\bibname{References}
\newcommand{\Msun}{\ensuremath{\rm \,M_\odot}}

\newcommand{\msun}{\ensuremath{\rm \,M_\odot}}

\newcommand{\cm}{\ensuremath{\,\mathrm{cm}}}

\newcommand{\kms}{\ensuremath{\,\mathrm{km}\,\mathrm{s}^{-1}}}
\newcommand{\gcm}{\ensuremath{\,\mathrm{g}\,\mathrm{cm}^{-3}}}
\newcommand{\ergs}{\ensuremath{\,\mathrm{erg}\,\mathrm{s}^{-1}}}
\newcommand{\gms}{\ensuremath{\,\mathrm{g}\,\mathrm{s}^{-1}}}
\newcommand{\msy}{\ensuremath{\Msun\mathrm{\; yr}^{-1}}}

\newcommand{\Ledd}{\ensuremath{\,L_{\rm Edd}}}
\newcommand{\Mdoted}{\ensuremath{\,\dot M_{\rm Edd}}}

\newcommand{\Teff}{\ensuremath{\,T_{\rm eff}}}

\begin{document}

\thispagestyle{empty}

\begin{center}
\vspace{0.5cm}

%\LARGE{\bf CHAPTER}

\setcounter{chapter}{1}

\vspace{0.5cm}
\Large{\bf Problems in the astrophysics of accretion onto compact celestial bodies
$^*$\blfootnote{$^*$Invited book chapter in "Highlights of the  Compact Objects Sciences in the Last Decade", edited by \c{S}\"olen Balman, IU Press (Istanbul University Press, Turkey)}} 

\vspace{1.0cm}

Jean-Pierre LASOTA$^{1,2}$\\

\vspace{1.0cm}
\small{
$^1$ Nicolaus Copernicus Astronomical Center, Polish Academy of Sciences,
           Warsaw, Poland\\ 
\vspace{0.25cm}
$^2$ Institut d'Astrophysique de Paris, CNRS et Sorbonne Universit\'e,
           Paris, France\\
E-mail: lasota@iap.fr\\
}
\end{center}

\hfill {\footnotesize{\sl Most people do not read; if they read, they do not understand. And those who understand forget.}}

\hfill                                {\scriptsize{Henry de Montherlant (French writer; 1895-1972)}}

\vspace{1.0cm}

\noindent
{\bf  Abstract}
Although during the last decade new observations and new theoretical results have brought better understanding
of the physics of accretion onto compact objects, many old and several new questions and problems await answers
and solutions. I show how the disc thermal--viscous instability model applied to both cataclysmic variable stars
and X--ray binary transients compels us to conclude that assuming the existence in these systems of a flat
accretion disc extending down to the accretor's surface or to the last stable orbit and fed with matter at its outer
edge is too simple and inadequate a description of these objects. It is also clear that, in most cases, these discs cannot driven
by (anomalous) viscosity only. The origin of the superhumps observed in cataclysmic variables and X--ray binaries
is, contrary to the common opinion, still unknown. In accreting magnetic white dwarf systems outbursts not of the dwarf--nova type 
can be due to the magnetic gating instability and/or thermonuclear micronova explosions. Although the ``typical''
lightcurves of X--ray transients can be described by analytical formul{\ae} (but their decay phase is {\sl not} exponential),
observations show that in many cases the light variations in these systems are much more complex.
An elementary argument shows the impossibility of magnetars in pulsing ultraluminous X--ray systems, but we still do not have a complete,
self-consistent description of supercritical accretion onto magnetized neutron stars and the resulting (necessarily beamed)
emission. Although it is (almost) universally believed that active galactic nuclei contain
accretion discs of the same type as those observed in binary systems, the evidence supporting this alleged truth is
slim and the structure of accretion flows onto supermassive black holes is still to be determined.
{\scriptsize 
\noindent 

\vspace{0.5cm}
}
\vspace{1.0cm}

\section{Introduction}
\label{sec:intro}

\subsection{Accretors}

There are three types of compact celestial bodies in the Universe: white dwarfs, neutron stars and black holes. Other types of compact objects have also been
proposed, e.g. strange stars, gravastars, Q-stars, boson stars etc., but so far, no evidence of their existence has been found. 
Compactness is defined through the size of the body relative to its gravitational radius $R \sim R_g = GM/c^2$. For white dwarfs, this is $R_g/R \sim 10^{-3} - 10^{-4}$,
so they are ``weakly'' compact, but their efficiency of accretion $\sim R_g/R$ can be comparable to the efficiency of thermonuclear reactions. Neutron stars are obviously
relativistic bodies with $R_g/R \sim 0.1$.

Black holes are purely general--relativistic objects. However, this does not mean that they must be extremely dense, or that the
gravitation at their surface must be extremely strong, contrary to what is (too) often asserted in the media and even in the astrophysical literature.
The mean density of a non--rotating black hole ($R_{\rm BH}=2R_g$) is equal to
\begin{equation}
\label{eq:densBH}
\rho_{\rm BH}(M)= \frac{3M}{8\pi R_g^3}= 1.0 \left(\frac{1.35\times 10^8}{M/\Msun}\right)^{2}\, \gcm,
\end{equation}
({\sl decreasing} with the square of the black hole mass) so the mean density of a $4 \times 10^9 \Msun$ black hole is the same as the density of air. It's not clear what the density of a black hole corresponds to, but since the space--time curvature is also $\sim M/R^3$, Eq. (\ref{eq:densBH}) tells us that the curvature near a black-hole horizon is not necessarily strong. The space-time curvature in the vicinity of the famous M87* black hole ($M=6.5 \times 10^9 \,\Msun$) is weaker than on the Earth's surface. In this sense, gravitation is stronger around us than it is near this six--billion solar-mass compact body.
But this does not exempt us from using Einstein's theory of gravitation when describing physical processes in the vicinity of a supermassive black hole. Although the pull of gravitation can be {\sl locally} suppressed by free--falling, crossing a black--hole surface, despite being unnoticeable, has literally inescapable consequences. Above the black--hole's surface, its gravity pull can be locally counteracted by applying external forces (through engines) but at the surface itself this requires infinite energy.
What is, usually, called the ``black--hole surface'' is, in fact a 2D slice of a stationary and axisymmetric null hypersurface -- a global, space--time structure. This has consequences that are, too often, ignored by astrophysicists.

The space--time curvature certainly cannot be ignored when considering motions on the scale of the curvature radius, such as propagation of light emitted near the black hole surface
(see e.g. \citealt{Gralla0719,Vincent0221}). Also the popular ``pseudo--potential'' of \citet{Paczynsky0880} should be used with care, since it describes correctly only Keplerian orbits (in the Schwarzschild metric) of massive particles in the equatorial plane, but not other types of motions (and not the light propagation, of course).

Black holes supposedly come in three mass categories: stellar--mass, intermediate--mass and supermassive (sometimes just called ``massive''). The first category is observed mostly in binary systems with a normal star, but since 2015, mergers of couples of such black holes have been observed in gravitational waves. While the masses of black holes in electromagnetically observed binaries span the range $ 5 - 21 \Msun$ \citep{millerJ0321}, in the case of detection through gravitational waves this range is much larger, spanning $\sim 3 - 80\Msun$ \citep{Abbott0123}. This difference could be due to the fact that observations in gravitational waves reaching up to $z \lesssim 0.7$, access a much larger range of metallicities than observations in electromagnetic waves \citep{Belczynski1121}. Low metallicities allow the formation of high--mass ($ > 30\Msun$) black holes \citep{Belczynski0610}. The existence of intermediate--mass black holes (IMBHs) is still to be confirmed. It is now clear that they are not components of the ultraluminous X--ray sources \citep[see, e.g.,][]{king0623}, but black hole mergers can result in $\gtrsim 100\Msun$ objects, formally in the IMBH range. The existence of supermassive black holes in galaxy centres is well established. The maximum mass of an accreting black hole could be equal to $2.7 \times 10^{11}\Msun$ \citep{King0216}, but this result relies on the (untested, see Sect. \ref{sec:AGN}) hypothesis that standard accretion discs are present in active galactic nuclei (AGN). It is conceivable that primordial, ``stupendously'' massive ($10^{12}- 10^{18}\, \mathrm{ M_{\odot}}$) black holes exist \citep{Carr02221}.

\subsubsection{Accretion flows}

Most of accretion flows onto compact celestial bodies contain a sufficient amount of angular momentum to form a flattened rotating structure, usually called an ``accretion disc'' even when it looks more like a torus. In binary systems, an accretion disc always form when the mass--loosing star is filling, or almost filling, its Roche lobe. It is not clear what is the structure of the inner accretion flow in active galactic nuclei (see e.g. \citealt{Antonucci0313} and Sect. \ref{sec:AGN}), but in at least some cases a spectacular disc is observed far from the nuclear black hole (e.g. a warped Keplerian disc in NGC 4258; \citealt{Miyoshi0195}). 

In the case of cataclysmic variables (CVs), thanks to observations of eclipsing systems, there is no doubt that the accreting white dwarf is surrounded by a geometrically thin Keplerian disc. Some such discs have been observed in a bright steady state (e.g., \citealt{Horne1085}), other in a non--steady quiescent state (e.g. \citealt{Wood0486}). Despite this certainty, the predictions of accretion disc models do not always correspond to observations. This is the case of disc spectra which fail to be faithfully reproduced by the standard $\alpha$--disc model (see e.g. \citealt{Idan0910}). The problem lies in the disc's vertical structure which still not well understood \citep{Hubeny0621}. In other systems, in cases where the existence of an accretion disc is not well established, model predictions not corresponding to the observed spectra (as e.g. \citealt{Kinney0189}) should not be automatically  used as a decisive argument against the disc's presence.

It seemed that the discovery by \citet{Balbus0791} that the magneto--rotational--instability (MRI) triggers turbulence in Keplerian discs, had finally solved the problem
of the origin of the accretion driving mechanism, but in reality things are more complicated. First, the MRI works only in ionised discs, while quiescent (low temperature)
discs between dwarf--nova and transient X--ray binary outbursts are neutral, hence not subject to MRI \citep{Scepi0118}. The same is true of protostellar accretion discs \citep{Lesur0221}. Therefore some other angular--momentum transport mechanisms have to be at work in such discs (e.g., winds, \citealt{Scepi0619}). Second, in hot discs of dwarf
nov{\ae} and X--transients in outburst, the observed decay times imply a viscosity parameter $\alpha$ (corresponding to the ratio of the (vertically averaged) total
stress to thermal (vertically averaged) pressure) $\approx 0.2$ (in the case of dwarf nov{\ae}, \citealt{Kotko0912}) and $\geq 0.2$ (for X--ray transients, \citealt{Tetarenko0218}), while MRI simulations (with no net magnetic field) give an $\alpha \sim 0.01$. In addition, the disc instability model (DIM) that correctly reproduces the main properties of dwarf--nova and X-ray transient outbursts requires a ratio 4 -- 10 between the values of $\alpha$ in the hot and cold states of the disc \citep{Smak0184,Meyer0384,Hameury0898}. This is not reproduced by the standard MRI simulations. $\alpha$ increases to $\sim 0.1$ when convection appears in the MRI simulations (\citealt{Hirose0514}) but this is not sufficient to produce dwarf--nova lightcurves resembling the ones observed \citep{Coleman116}. It seems that additional ingredients, such as winds (\citealt{Scepi0619,Tetarenko0218}) must play a role in driving disc accretion.

Despite its obvious weaknesses, the DIM (see \citealt{Lasota0106,Hameury0920} for reviews of the model) has proved to be a powerful tool to test some basic properties of accretion discs, mainly those relating to the accretion driving mechanisms. The best test-bed for the MRI simulations are CV discs, in particular those of dwarf--nov{\ae}, since they are the real structures closest to what these simulations are supposed to describe. Due to a couple incorrect determinations of the distance to the closest dwarf nova SS Cyg, the veracity of DIM was put out to doubt, but its reputation has been rescued by distance measurements by radio interferometry and Gaia (see \citealt{Schreiber0513} and references therein).

In this chapter I present and discuss problems with understanding accretion onto compact objects, that have been solved or have arisen mainly during the preceding decade. As usual, solving some problems, gives rise to new ones. The chapter begins with a reminder of the basic DIM features. This is followed by a discussion of the problems of applying this model to the description of the observed dwarf--nova outbursts. I then present the inadequacy of the almost universally used model of superhumps but I also enumerate the weaknesses of the alternative. The next section deals with outbursts of systems with magnetised white dwarfs. I consider first the (rare) cases when such binaries appear as dwarf nov{\ae}, then I present the successful application of the magnetic--gating instability model, designed to explain the neutron--star Rapid Burster, to the case of intermediate polars. I end this section with a discussion the problems with the recently proposed phenomenon of micronov{\ae}. The following part of the article deals with X-ray binaries. I begin with presenting an analytic method of describing the decay--from--maximum light curves of X-ray transients. Then I provide a detailed discussion of pulsing ultra--luminous X-ray sources (PULXs) in which I present a simple argument that exclude the presence of magnetars in these systems. A short subsection deals with recent results on transient ULXs. The chapter ends with a section concerned with the problem of accretion discs in AGNs.

\section{The thermal--viscous disc--instability}\label{sec:DIM}

If they are sufficiently large, all hot ($T > 10^4$K), standard, stationary accretion discs are thermally unstable.
In such discs the viscous heating rate per unit surface can be written as
\begin{equation}
\label{eq:vischeat2}
Q^+=\frac{{\mathfrak T\Omega^{\prime}}}{4\pi R}=\frac{9}{8} \Sigma \nu \Omega_K^2,
\end{equation}
where $\Omega$ is the angular velocity, the prime denotes the radial derivative, $\Sigma=\int^{+\infty}_{-\infty}\rho\,dz$ is the column density and $\mathfrak T$ the total ``viscous" torque (see e.g. \citealt{Lasota16}).
In the last equality, the disc is assumed to be Keplerian ($\Omega=\Omega_K$).

For Keplerian discs the angular momentum conservation in the form of
\begin{equation}
\dot M (\ell -\ell_{\rm in}) ={\mathfrak T},
\end{equation}
$\ell$ being the specific (per unit mass) angular momentum (the subscript ``in'' designs the its value at the flow's inner limit),
implies the following relation between viscosity and accretion rate
\begin{equation}
\label{eq:amintK}
\nu \Sigma =\frac{\dot M }{3\pi}\left[1 -\left(\frac{R_{\rm in}}{R}\right)^{1/2} \right],
\end{equation}
where ${R_{\rm in}}$ is the inner disc radius.
Therefore, from Eqs. (\ref{eq:vischeat2}) and (\ref{eq:amintK}) it follows that
\begin{equation}
\label{eq:teff}
Q^+ \equiv \sigma T_{\rm eff}^4= \frac{3}{8\pi}\frac{GM\dot M}{R^3}\left[1 -\left(\frac{R_{\rm in}}{R}\right)^{1/2} \right],
\end{equation}
hence the temperature of the disc is {\sl decreasing} with distance from the centre:
\begin{equation}
\label{eq:temprofile}
T_{\rm eff}\sim R^{-3/4}.
\end{equation}
Notice that Eq. (\ref{eq:temprofile}) assumes only the stationarity and Keplerianity of the disc, so it is a universal relation independent of the accretion mechanism.

It follows that for $r=R/R_S \gg 1$ ($R_S\equiv 2R_g$) the temperature profile of a stationary Keplerian accretion disc can be written
\begin{equation}
\label{eq:Teff_value}
T_{\rm eff}=T_{\rm in} \left(\frac{r}{3}\right)^{-3/4},
\end{equation}
where
\begin{equation}
T_{\rm in}=\left(\frac{3GM\dot M}{8\pi \sigma (3R_S)^3}\right)^{1/4} \approx 3.0 \times 10^9\,  m^{-1/2} \dot m^{1/4}\rm K,
\label{eq:Teff_in}
\end{equation}
if one assumes that $R_{\rm in}=3R_S$, i.e. that it is the ISCO for a non-rotating black hole. $M=m \Msun$ and $\dot M=\dot m M_{\rm Edd}$, where the Eddington accretion rate
is defined as 
\begin{equation}
\Mdoted \equiv \frac{4\pi GM}{\eta \kappa_{\rm es}c} = 1.4\times 10^{18}\eta_{0.1}^{-1} \left(\frac{M}{\Msun}\right)\,{\rm g\,s^{-1}}, \\
\end{equation}
where $\eta=0.1\eta_{0.1}$ is the radiative efficiency of accretion and $\kappa_{\rm es}$ the electron--scattering (Thomson) opacity coefficient.

For white dwarf accretors the ISCO is not relevant since the stellar radius $R_\star \gg R_S$ and typical accretion rates are usually well below the Eddington value.
In this case the disc temperature profile is more conveniently written as
\begin{equation}
\label{eq:Teff_valueWD}
T_{\rm eff}=T_{\rm in} \left(\frac{R}{R_\star}\right)^{-3/4},
\end{equation}
with 
\begin{equation}
T_{\rm in}=\left(\frac{3GM\dot M}{8\pi \sigma R_\star^3}\right)^{1/4} \approx 4.1 \times 10^4\,  m^{1/4} \dot M_{16}^{1/4}R_9^{3/4}\rm K,
\label{eq:Teff_inWD}
\end{equation}
where $R=R_910^9$cm.

%\begin{wrapfigure}{r}{0.55\textwidth}
\begin{figure}
  \centering
    \includegraphics[trim=0 0 0 0, clip, width=0.8\textwidth]{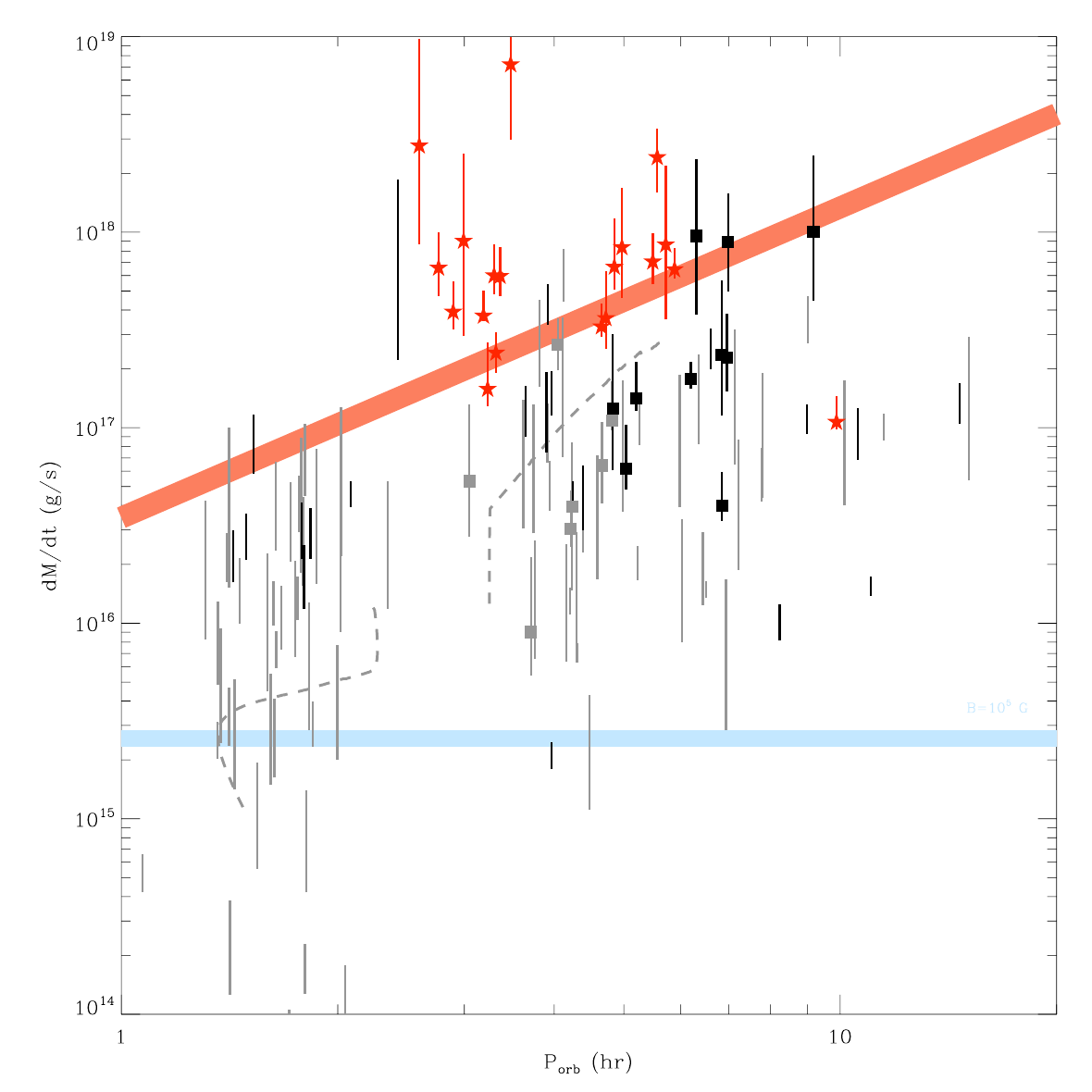}
  \caption{Mass transfer rates of CVs compared to the stability criterion. Systems above the upper (red) solid line are hot and stable. Systems below
the lower (blue) line indicate cold, stable disks if the white dwarf magnetic field $B \geq 10^5$ G. The dashed line represents the expected secular mass
transfer rate \citep{Knigge0611}. Square symbols indicate Z Cam type dwarf nov{\ae}; (red) stars indicate nova-likes. Dwarf nov{\ae}
shown in black have a more complete observed light-curve than those in grey. ({\sl From \citealt{Dubus0918}}.)}
\label{fig:cvstability}
%\end{wrapfigure}
\end{figure}
We see that, for sufficiently large discs, depending on the accretor's mass and accretion rate, even for very hot discs, a radius will be reached where the temperature drops below $\sim 10^4$K, roughly the hydrogen recombination temperature. This is where the disc not only stops being hot but also becomes thermally unstable (see, e.g. \citealt{Hameury0920}). For neutron stars and stellar--mass black holes, the critical (maximal) hot--disc radius is $10^{11} - 10^{12}$cm for $\dot m \sim 1$ and 
$\gtrsim 10^{10}$cm for $\dot m \approx 10^{-3}$. For white dwarfs, this radius varies from $\sim 10^{10}$cm for $\dot M_{16} \lesssim 1$, to $\gtrsim 10^{11}$cm
for $\dot M_{16} \gtrsim 100$ (notice that such accretion rates do not result in luminosities close to the Eddington value, because the accretion efficiency $R_g/R_*$ for white dwarfs is $\lesssim 0.001$ and not $\sim 0.1$).

Therefore cataclysmic variables, whose accretion discs have radii larger than the critical value, cannot be steady. 

\section{Cataclysmic variables}
\label{sec:CV}

\subsection{Dwarf nov{\ae}}
\label{sec:DN}

Figure \ref{fig:cvstability} shows that they do indeed exhibit outbursts: they are dwarf-nova stars which show repeated outbursts with
an amplitude larger than about 2 optical magnitudes on timescales of weeks to decades. In this figure the solid red line corresponds to the stability criterion obtained from Eq. (\ref{eq:Teff_valueWD}) by assuming $\Teff(R)=T_{\rm crit}$, where $T_{\rm crit}$ is the value of the temperature at which the disc becomes thermally unstable. The critical values of parameters at which the disc becomes unstable are calculated through fits obtained from numerical models of the disc's vertical structure. The stability limit in Figure \ref{fig:cvstability} uses fits from \citet{Lasota0808} where $\Teff(\rm crit)=6890$K. Eq. (\ref{eq:Teff_valueWD}) provides $\dot M_{\rm crit}(R_\mathrm{D})$, so to get $\dot M_{\rm crit}(P_{\rm orb})$ one uses a relation between the disc radius $R_\mathrm{D}$ and the orbital period $P_{\rm orb}$, obtained assuming that the radius of the disc is a fraction $f(q)$ of the binary separation $a$:
$R_{\mathrm{D}}= f(q)a=2.28 \times 10^9 f(q) M_1^{1 / 3} P_{\min }^{2 / 3} \mathrm{~cm}$, where $q$ is the mass ratio (mass-of-the-companion/white-dwarf-mass). In general, $f$ is well approximated by $f=0.6/(1+q)^{2/3}$.

In Fig. \ref{fig:cvstability} all the observed systems above the stability limit are steady (``nova--like''), as they should be; all systems below show outbursts, except for one: a very special magnetic binary AE Aqr, also known as a source of very--high energy emission, for which the method of deducing the mass--transfer rate from the secondary used in plotting the figure does not apply (see \citealt{Dubus0918} for details). 

\subsection{Dwarf--nova lightcurves}
\label{sec:dnlc}

The DIM for dwarf nov{\ae} is not only successful in predicting which CV must be a dwarf nova, but also is able to reproduce the lightcurves of (at least some) of these systems.
\begin{figure}[ht]
    \centering
    \includegraphics[width=0.5\textwidth]{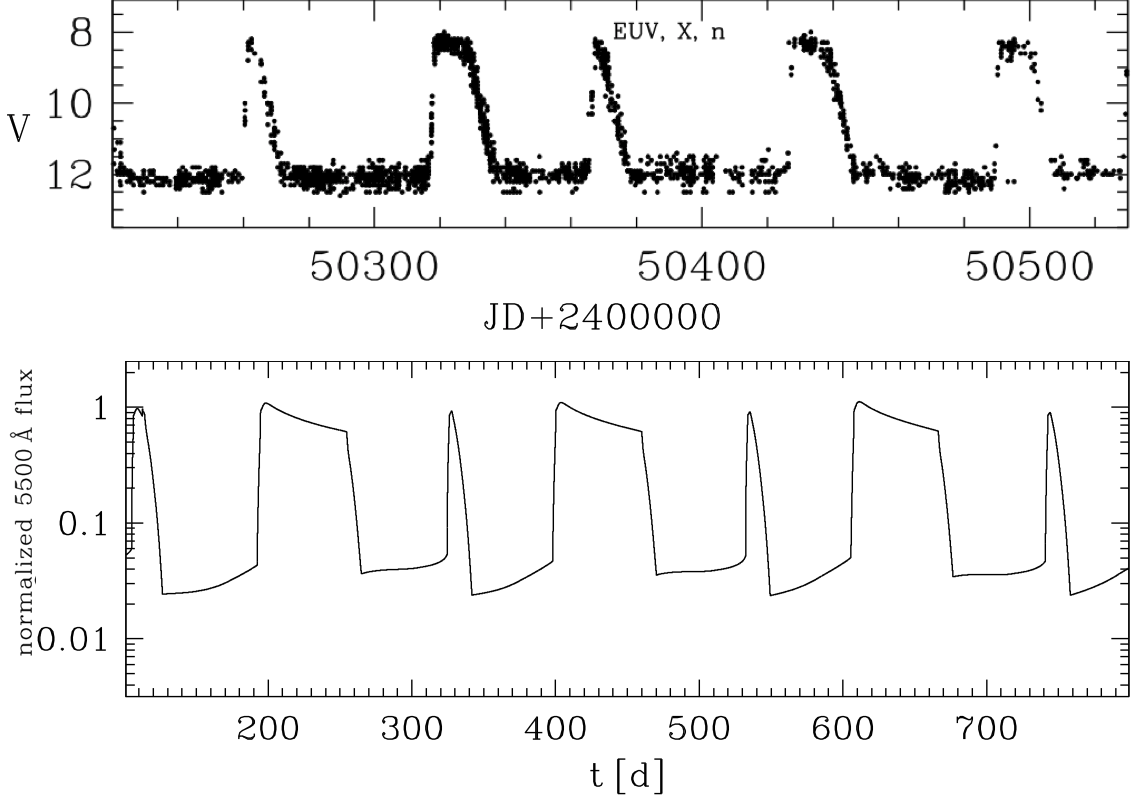}
    \caption{Comparison of the observed light curve of the dwarf nova SS Cyg (top; data from the {\it Association Fran\c caise des Observateurs d'\'Etoiles Variables}), with lightcurve calculated form DIM 
    for a system with the same as SS Cyg parameters (bottom). ({\sl Adapted from \citealt{Schreiber1003}}).}
    \label{fig:sscyglc}
\end{figure}
Figure \ref{fig:sscyglc} \citep{Schreiber1003} compares observations of SS Cyg with the lightcurve calculated using the DIM code by \citet{Hameury0898}.
The DIM reproduces the two types of outbursts observed in this best--observed (and brightest) dwarf nova, as well as the
recurrence time. However, one should stress that the ``standard'' DIM cannot reproduce the observed sequences of dwarf--nova outbursts (see e.g. \citealt{Hameury0920,Lasota0106}).
By ``standard'' (or ``basic'') I understand the model which assumes that the disc always extends to the white-dwarf surface and is supplied in mass to its outer edge at a  constant rate. To calculate the lightcurve in Fig. \ref{fig:sscyglc}, the disc was assumed to be truncated and the mass--supply was smoothly varied by $15\%$.
However, SS Cyg, a dwarf-nova of U Gem type, has a relatively simple lightcurve, if one exclude occasional anomalies. 

In other type of systems the lightcurves are more complex. For example, in Z Cam--type dwarf nov{\ae}, the decay from outburst's peak is interrupted by a standstill. In this case the DIM can reproduce the lightcurves of Z Cam stars if one takes into account the heating of the outer disc by the impact of the mass--transfer stream and by the tidal torques and if the mass-transfer rate from the secondary varies by about 30\% around the value critical for stability \citep{Buat0401}. In this case, the disc during standstill is hot and stable. This, however, seemed to be false when outbursts appearing during standstills were observed \citep{Simonsen0611,Szkody1213}. \citet{Hameury0914} showed that applying the DIM to such systems requires a rather special type of mass--transfer bursts from the secondary. Such bursts should last a few days and have short rise--times and exponential decays  followed by short but significant mass-transfer dips.  They could result from a giant
flare near the Roche--lobe filling, secondary's star surface, due, for example to the absence of star spots in the L1 region.

All these truncations and mass--transfer variations could look like made--up tricks had they not been observed in real CVs. Disc truncation is confirmed by X--ray observations (see e.g. \citealt{Balman0114,Dobrotka0423}) and could be due either to the action of the white--dwarf's magnetic field (see Sect. \ref{sec:IPs}) or to evaporation \citep{Meyer0894}. Huge mass--transfer variations are directly observed in polars (AM Her stars; see \citealt{Kalomei0512} and references therein). Since the strong magnetic moment of the white dwarf prevents disc formation in these systems, there can be no doubt that the observed luminosity variations are provoked by changes in the mass--transfer rate from the companion star. Also in the case of VY~Scl (see Sect. \ref{sec:IPs}) it is clear that the mass-transfer simply switches off. Short--term variations are most probably provoked by the movements of star spots, as other mechanisms involving the star's bulk, or the mass transfer  are excluded \citep{Ritter0888}.
\begin{figure}[ht]
    \centering
    \includegraphics[width=0.9\textwidth]{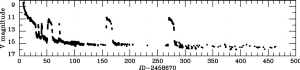}
    \caption{Lightcurve of the SU UMa system TCP J21040470+4631129. Data from AAVSO.}
    \label{fig:TCP}
\end{figure}

Things become even more complicated when one tries to apply the DIM to dwarf--nov{\ae} called SU~UMa stars. In these systems, in addition to normal
dwarf--nova outbursts, one also observes longer and brighter eruptions, so-called superoutbursts. Superoutbursts occur regularly, typically every few
normal outbursts. A sub--class of SU~UMa stars, the WZ~Sge--type dwarf nov{\ae}, show superoutbursts only. The nature of the mechanism producing superoutbursts
is still a subject of controversy. On the one hand, the tidal--thermal instability (TTI) model, proposed by \citet{Osaki0189}
invokes the existence of a tidal instability that is supposed to occur when the mass ratio is low enough for
the 3:1 resonance radius to fit inside the tidal radius defining the outermost disc radius (see Sec. \ref{sec:shmp} for details and the allegedly related superhump phenomenon). 
This tidal instability is supposed to generate an increased viscous dissipation in the whole disc, thus leading to a superoutburst until
the disc shrinks enough to be well inside the tidal radius. The physical mechanism behind the supposed viscosity--increase mechanism in
TTI remains a total mystery.

On the other hand, \citet{Vogt0283}, \citet{Smak0191} and \citet{Osaki0385} proposed that superoutbursts are caused by an enhanced mass transfer (EMT) from the secondary. Next, \citet{Hameury0100}  showed that relating the mass transfer rate to the accretion rate, i.e. assuming that irradiation of secondary increases the mass--transfer rate, allows one to reproduce the observed visual light curves. The irradiated EMT (IEMT) works quite well (see e.g. \citealt{Schreiber1104}) and succeeds where the TTI model fails, i.e. it reproduces the lightcurves of the ER UMa's, those frantic SU UMa-type stars, that are never in quiescence, showing superhumps (see next Section) even when exhibiting normal outbursts.
However, while in contrast to the TTI, the IEMT superoutburst mechanism is well--specified, it is not immediately clear that it can work in real binaries. The problem is that, because it is shadowed by the disc, the L1 point cannot be irradiated directly, which precludes any significant increase of the mass transfer rate \citep{Viallet1107}. This is true if the disc is flat. But discs in binary systems might be warped which would allow the L1 point to be irradiated directly, at least at some orbital phases. Simulations by \citet{Cambier1015} show that when the disc is warped, irradiation can ``agitate'' the mass--transfer from the secondary. The simulations leading to this conclusion are still only in 2D, we shall have to wait some time for a detailed description of the irradiated mass--transfer enhancement. As we shall see in a moment, a warped disc has another useful property: it allows the mass--transfer stream to over(under)flow the disc surface, modifying, among other things, how mass and angular momentum are delivered to the disc. One should stress, however, that in cataclysmic variables the warping mechanism is not well established. In fact, the only viable mechanism proposed is itself related to secondary's irradiation. \citet{Smak1209} finds that the mass--transfer stream through L1 has a component perpendicular to the disc plane which oscillates in phase with the binary period. He suggests that this comes about because the tilted disc enables the neighbourhood of the L1 point to be heated in an asymmetric manner, which varies with the orbital period.
\begin{figure}[ht]
    \centering
    \includegraphics[width=0.4\textwidth]{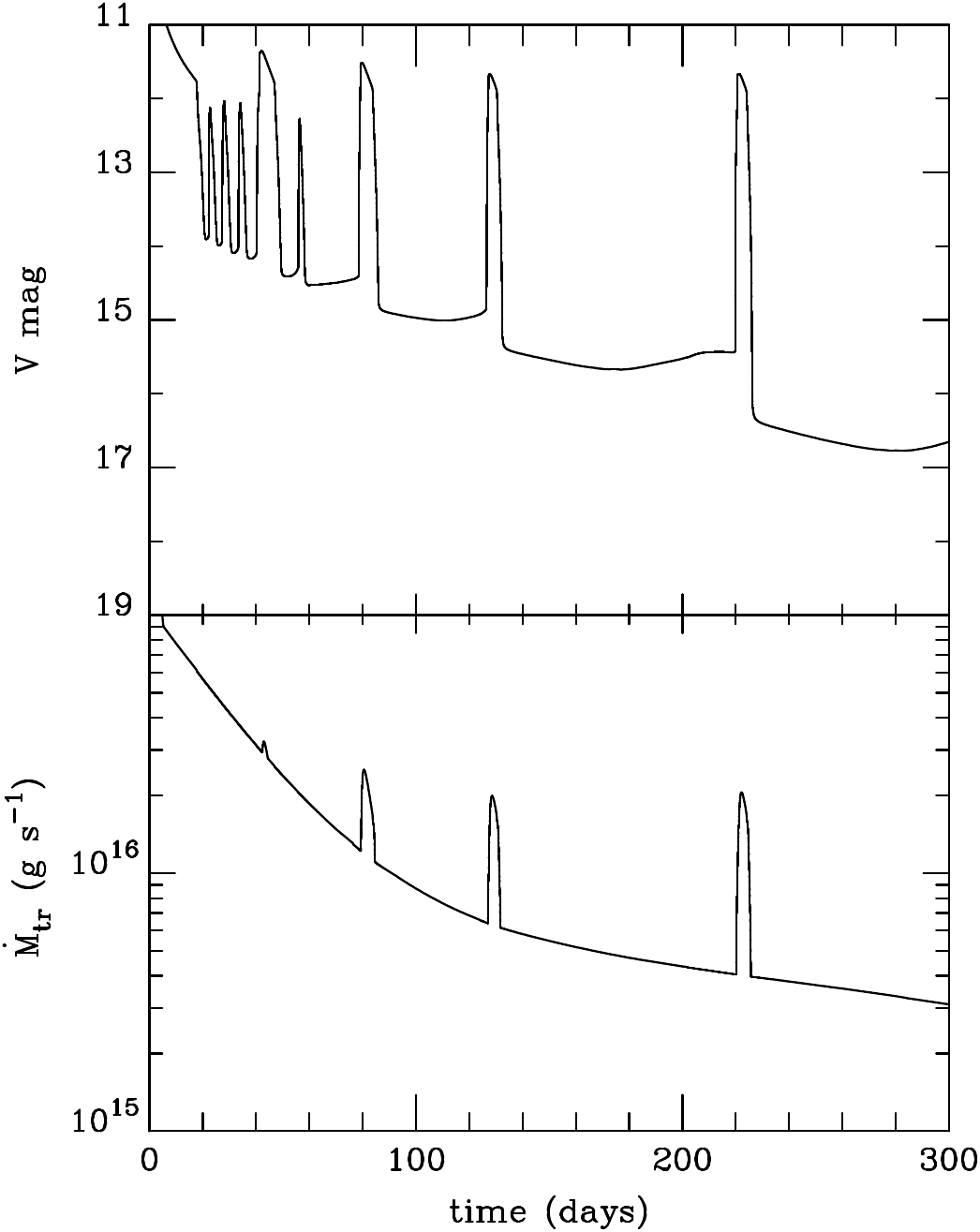}
    \caption{Calculated lightcurve similar to that observed in TCP J21040470+4631129.
Top panel: V magnitude, including contributions from the white dwarf
and the hot spot. Bottom panel: mass transfer rate from the secondary. ({\sl From \citet{Hameury0621}.})}
    \label{fig:tcplc}
\end{figure}

Reproducing the extraordinary lightcurve of the dwarf nova TCP J21040470+4631129 (Fig. \ref{fig:TCP}) with the DIM is a real challenge. This system exhibited, first a bright superoutburst, followed by two normal outbursts that were succeeded by three superoutbursts (fainter than the first one), the first two of which were separated by a normal outburst.
In total 4 superoutbursts and 3 normal outbursts during 300 days. All superoutbursts exhibited superhumps.

\citet{Hameury0621} showed that the main features of this astonishing lightcurve can be reproduced by the DIM with the following additions: the mass transfer rate from the secondary increases by several orders of magnitudes during the
initial superoutburst. Then the mass--transfer rate slowly returns to its secular average and causes the observed succession of outbursts with
increasing quiescence durations until the disc is steady, cold, and neutral; its inner parts are truncated either by the white dwarf
magnetic field or by evaporation. The very short, quiescence phases between reflares are reproduced when the mass-transfer stream
overflows the disc. The luminosity in quiescence is dominated by a hot white dwarf that cools down on timescales of months.

Using similar additions to the DIM, one can also produce lightcurves containing rebrightenings closely resembling those observed in two WZ~Sge stars,
the prototype and EG Cnc \citep{Hameury0621}. 

All these supplements may look like ``epicycles''\footnote{On the other hand, epicycles, deferents and equants do not deserve their bad name (see \href{https://farside.ph.utexas.edu/books/Syntaxis/Almagest/index.html}{A Modern Almagest})}, 
but their necessity proves that an accretion
disc by itself, even when a realistic model is used to describe its behaviour, will not be able to reproduce lightcurves of TCP J21040470+4631129's complexity.

\subsection{The superhump problem}
\label{sec:shmp}

Superhumps are periodic light variations, with periods slightly longer (in the case of {\sl positive} superhumps\footnote{There exist also negative superhumps, with slightly shorter periods.}) than the
orbital period, observed in the light--curves of dwarf--nova superoubursts.
Superhumps are also observed in some bright (nova--like) cataclysmic variables, the so--called ``permanent superhumpers'' (\citealt{Patterson0199}).

A popular explanation of the superhump phenomenon is given by the 
tidal-resonance model (TR model, Whitehurst 1988, Hirose and Osaki 1990) according to which 
it results from 
periodic enhancement of tidal stresses in an eccentric accretion disc undergoing
apsidal motion. The mechanism producing the disc's eccentricity is supposed to be 
provided by the 3:1 resonance between the orbital frequency of the binary system and 
the orbital frequency of the outer parts of the deformed disc. For this mechanism to work,
the 3:1--resonance radius
\begin{equation}
\label{eq:31rad}
R_{3:1} =\frac{1}{3^{2/3}(1+q)^{1/3}  }a, 
\end{equation}
where $a$ is binary separation, must be smaller than the disc (tidal) radius.
For a long time and not very--well known reasons it was believed that the value of the maximum mass--ratio 
for the $R_{3:1}<R_{\rm tid}$ condition to be satisfied is  $q_{\rm crit}=0.25$ or even 
$q_{\rm crit}=0.39$. \citet{Smak1220a} calculated this critical ratio from the first principles obtaining a smaller value: $q_{\rm crit}=0.22$.

All observed permanent superhumpers have $q> 0.24$. The prototypical dwarf--nova U~Gem have $q=0.35$, but in 1985 went into a superoutburst. In dwarf nov{\ae} of the SU~UMa type, superhumps appear during superoubursts, so when
\citet{Lasota1095} have pointed out that the outburst of U~Gem observed in X--rays by \citet{Mason0688} is in fact
a superoutburst (the only one seen during almost 170 years of continuous observations), \citet{Smak1204} searched in the archival data for a superhump
and found one. Its statistical significance is only $2\sigma$ \citep{Schreiber0507} but if superhumps in dwarf nov{\ae}
are related to superoutbursts, its presence should be expected. The argument against its reality on account of the
``incorrect'' mass--ratio of the system, has been considerably weakened by the observation of equally incorrect permanent superhumpers.
But there is worse.

The frequency of the eccentric--disc apsidal motion $\Omega_{\rm abs}$ is related to the orbital and superhump frequencies $\Omega_{\rm orb}$ and $\Omega_{\rm SH}$ through
\begin{equation}
\Omega_{\rm abs} = \Omega_{\rm orb} - \Omega_{\rm SH}.
\end{equation}
The superhump excess, used to quantify the superhump phenomenon is defined as
\begin{equation}
    \varepsilon_{\rm SH}\equiv \frac{P_{\rm SH} - P_{\rm orb}}{P_{\rm orb}},
\end{equation}
($P=2\pi/\Omega$) and can be expressed through the apsidal frequency:
\begin{equation}
    \varepsilon_{\rm SH}=\cfrac{\Omega_{\rm aps}}{\Omega_{\rm orb} - \Omega_{\rm aps}}.
\end{equation}
The initial version of the TR model assumed that the apsidal motion can be described by orbits of free particles and that
this dynamical effect is given by a function of the mass ratio and the disc's effective radius $\Omega_{\rm dyn}=f(q,R)\,\Omega_{\rm orb}$,
where the effective radius is assumed to be the 3:1 resonance radius in Eq. (\ref{eq:31rad}).
This formulation of the model failed, however, to describe the observations (see e.g. \citealt{Murray0500,Pearson0906}). 
But since, in reality, particles in the discs in question do not move on exactly free orbits, one could hope that adding a pressure term to the
apsidal frequency formula:
\begin{equation}
    \Omega_{\rm aps}= \Omega_{\rm dyn} + \Omega_{\rm press},
\end{equation}
would solve the problem
(\citealt{Murray0500,Montmogery08201,Pearson0906}).

\begin{figure}[ht]
    \centering
    \includegraphics[trim= 0 7cm 0 1cm, clip, width=0.8\textwidth]{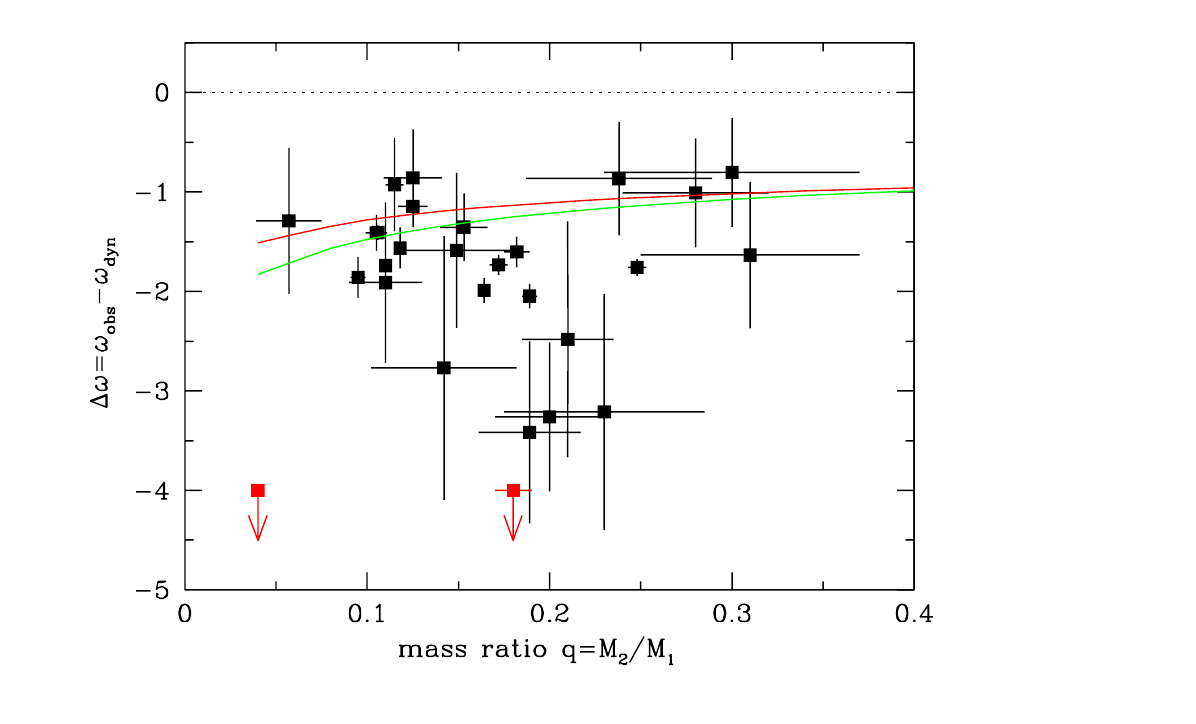}
    \caption{The residuals $\Delta \Omega$ (see text for details) plotted against the mass ratio. Red symbols
represent the two helium CVs. Red and green lines are theoretical $\Omega_{\rm press}= f(q)$ relations from \citet{Montmogery08201} and \citet{Pearson0906}
respectively. ({\sl Adapted from \citealt{Smak1220b}}).}
    \label{fig:omp}
\end{figure}

\citet{Smak1220b} have tested this hypothesis by comparing its predictions to observations of 21~CVs exhibiting superhumps (including two helium, AM CVn systems). His results are presented in Fig. \ref{fig:omp}.
\citet{Smak1220b} has determined the observed apsidal frequency, then calculated $\Omega_{\rm dyn}$, using the system's orbital parameters. This allowed him to calculate $\Delta \Omega = \Omega_{\rm aps} - \Omega_{\rm dyn}$, which according to the theory should be equal to $\Omega_{\rm press}$ and plotted it as a function of $q$. Clearly, the result contradicts the model: not only do the points not follow the theoretical curves but they show large scatter and do not seem to represent any clear trend. Clearly, the  TR model is not in agreement with observation. Unfortunately, it is widely used for their interpretation.

The irradiation--modulated mass--transfer (IMMT) model (\citealt{Smak0309,Smak0917}), based on purely observational evidence, explains superhumps as being due
to the periodically variable dissipation of the kinetic energy of the stream which results from variations in the mass transfer rate which are produced by the modulated
irradiation of the secondary star. This is a purely ``observational model'', providing a description of the phenomenon but not its explanation. It does not provide
the mechanism of the clock ticking at the superhump period. Since the model assumes that the mass--transfer is modulated through the irradiation of the secondary, it would imply
the presence of a warped accretion disc, but the stream overflow of such a disc is supposed to produce only negative superhumps, not the excess--period ones observed. Of course the accretion stream can also overflow a
flat disc but then it is unlikely that the secondary's irradiation would be able to increase the mass--transfer rate. On the other hand, there are systems showing both negative
and positive superhumps, but they are rather an exception.

The superhump phenomenon in CVs still awaits a complete explanation (see e.g., \citealt{Oyang0721}).

\section{Instabilities in magnetic white--dwarf systems}
\label{sec:IPs}

When its moment is sufficiently large, the accretor's magnetic field can affect the structure of the inner accretion flow.
White dwarfs, whose dipolar magnetic fields can have more than $10^7$G, can have sometimes magnetic moments  larger than $10^{34} \rm  G\, cm^3$,
4 orders of magnitudes larger than that of the usual X-ray pulsars (XRPs) with magnetic fields $\sim 10^{12}$ G. Only the
most extreme magnetars have magnetic moments comparable to that of polars (AM Her stars), but since they are never members
of binary systems, this is not relevant for accretion processes.

The condition for accretion--disc formation and existence can be written as
\begin{equation}
    R_{\text {circ }} > R_{\mathrm{M}}
    \label{eq:rcrm}
\end{equation}
where the circularisation radius is defined through
\begin{equation}
    R_{\text {circ }}=3.5 \times 10^{10} m^{1/3} f(q) P_{\mathrm{orb}(\mathrm{h})}^{2 / 3} \mathrm{~cm},
\label{eq:Rcirc}    
\end{equation}
where $ 0.12 \lesssim f(q) \lesssim 0.3$, while the  magnetospheric radius can be written as
\begin{equation}
    R_{\mathrm{M}}=1.4 \times 10^{10} \dot{M}_{15}^{-2 / 7} m^{-1 / 7} \mu_{32}^{4/7} \mathrm{~cm},
\label{eq:RM} 
\end{equation}
where $\mu=10^{32} {\rm Gcm^3}\mu_{32}$ \citep{FKR2002}.

Therefore when $\mu \approx 10^{34}\rm G cm^3$, for systems with $P_{\mathrm{orb}} < 8$ hr, a disc will not form, even for $\dot{M}_{15} = 100$. 
This is case of polars, for which the magnetospheric radius is larger than the orbital separation and the secondary is brought by magnetic torques to 
rotation synchronous with the orbital motion \citep{Joss0579}. For lower magnetic moments, depending on the orbital period and mass--transfer rate, a disc might form,
as is the case of intermediate polars (IPs). However, if the mass--transfer from the companion star diminishes or even stops, such discs will vanish
when the increasing magnetospheric radius reaches and overtakes the circularisation radius. This phenomenon was used by \citet{Hameury1002} to explain
the absence of dwarf--nova outbursts during the decay and rise phases of the VY Scl stars luminosity variations. In these nova--like CVs, with orbital periods mostly between
3 and 4 hours, the mass--transfer diminished on timescales longer than the disc viscous time, reaching a minimum during which the transfer of mass stops.
Then on similar timescales the system reaches its initial, quasi--steady luminosity state. Since during the luminosity decay and rise the system crosses the
thermal--viscous instability strip, one would expect to observe dwarf--nova outbursts, but none has been observed. \citet{Hameury1002} found that this
can only happen if,  during the decay, the magnetospheric radius exceeds the circularisation radius, so that the disc disappears before it enters the instability strip for dwarf nova outbursts. And on the way up, the disc reappears only when it can be stable. The principle is simple: no disc outbursts must mean: no disc \citep{Hameury0605}.
But this must also mean that white dwarfs in VY Scl stars dwarf--nova outbursts are magnetised, with a magnetic moment
\begin{equation}
    \mu \gtrsim 1.5 \times 10^{33} f_{0.12}^{1.75} P_{\mathrm{orb}(\mathrm{4h})}^{2.06}\left(3 R_{\rm {out }} / a\right)^{1.34} m^{1.4} \mathrm{G}\, \mathrm{cm}^3,
\end{equation}
where $R_{\rm out}$ is the outer disc radius, and $a$ the orbital separation.

Polars and IPs exhibit low states similar to those observed in VY Scl stars. Since polars have no discs, the dimming of these sources can be due only to a drop of the mass--transfer from the secondary. But IPs with discs should be become discless by the same mechanism as VY Scl stars.
\citet{Hameury1017} showed that observations of the IP FO Aqr are well accounted for by the same mechanism that we have suggested to explain the absence of outbursts during low states of VY Scl stars. This has been confirmed in detail by the observations of this system by \citet{Littlefield0620}.

The hypothesis that all VY Scl stars contain magnetized white dwarfs still needs observational confirmation. Until now, only the VY Scl star DW Cnc has been confirmed to be an IP with a spin period $\sim 38.6$ min \citep{Rodriguez0304,Duffy0222}. Some other systems of this type could be magnetic according to e.g. \citet{Zemko1114}. Three SW Sex stars: LS Peg and V795 Her, RXJ1643.7+3402 are known to exhibit optical modulated circular polarization \citep{Rodriguez0509}, an unmistakable signal of the presence of a strong ($> 10^6$ G) magnetic field,

\subsection{Intermediate Polars as dwarf nov{\ae}}
\label{sec:IP}

Dwarf nova outbursts from intermediate polars are rare despite the fact that many of them having mass--transfer rates locating them in the
thermal instability strip, i.e. a mass--transfer rate satisfying
\begin{equation}
\label{eq:mdotplus}
    \dot{M}_{\mathrm{tr}}<\dot{M}^+_{\rm crit}\left(R_{\mathrm{out}}\right)=9.5 \times 10^{15} m^{-0.88}\left(\frac{R_{\mathrm{out}}}{10^{10} \mathrm{~cm}}\right)^{2.65} \mathrm{~g} \mathrm{~s}^{-1},
\end{equation}
where $R_{\mathrm{out}}$  is the ``effective'' disc radius (usually $\sim 0.8 R_D$; \citealt{Hameury0617}), and
\begin{equation}
\label{eq:mdotminus}
    \dot{M}_{\mathrm{tr}}>\dot{M}^-_{\rm crit}\left(R_{\mathrm{in}}\right)=8.4 \times 10^{12} m^{-0.89}\left(\frac{R_{\mathrm{in}}}{10^{9} \mathrm{~cm}}\right)^{2.68} \mathrm{~g} \mathrm{~s}^{-1}.
\end{equation}
In general, for hydrogen--dominated discs extending down to the white dwarf's surface, the lower limit is too low to be of much interest\footnote{But stable cold helium-dominated discs of AM CVn stars, satisfying $\dot{M}_{\mathrm{tr}}<\dot{M}^-_{\rm crit,He}\left(R_{\mathrm{in}}\right)$ are observed (see \citealt{Kotko0812}). Because of the higher ionisation potential of helium, $\dot{M}^-_{\rm crit,He} \gg \dot{M}^-_{\rm crit,H} $ \citep{Lasota0808}.}.

However, in the case of IPs, $R_{\mathrm{in}} \approx R_{\mathrm{M}}$, putting Eq. (\ref{eq:RM}) in Eq. (\ref{eq:mdotminus}) gives a stability condition\footnote{Here the effective inner radius is taken as $1.5R_{in}$ (see \citealt{{Hameury0617}})}
\begin{equation}
    \dot{M}_{\mathrm{tr}}< 6.6 \times 10^{15} m^{-0.72} \mu_{32}^{0.87} \mathrm{~g} \mathrm{~s}^{-1},
\end{equation}
which is quite a realistic mass--transfer rate for these systems.

IPs with no  disc ($ R_{\text {circ }} < R_{\mathrm{M}}$) obviously cannot have dwarf--nova outbursts. But the majority of IPs with magnetic fields allowing the presence of discs 
seem to be mostly steady, and the rare observed outbursts, in particular in systems with long orbital periods, are much too short (sometimes lasting less than the orbital period) to be dwarf--nova outbursts, since only long outbursts (lasting a few days) result from the thermal--viscous disc instability. In many cases the mass transfer is low enough and the magnetic field strong enough to keep the accretion disc stable on the cold equilibrium branch \citep{Hameury0617}.

\subsection{Magnetic--gating instability}
\label{sec:MGI}

The nature of the short (and rare) IP outbursts therefore requires an explanation, especially now that the Transiting Exoplanet Survey Satellite (TESS), with its unprecedented monitoring of the optical sky has drastically increased their observed number. 

\begin{figure}[ht]
    \centering
    \includegraphics[width=0.9\textwidth]{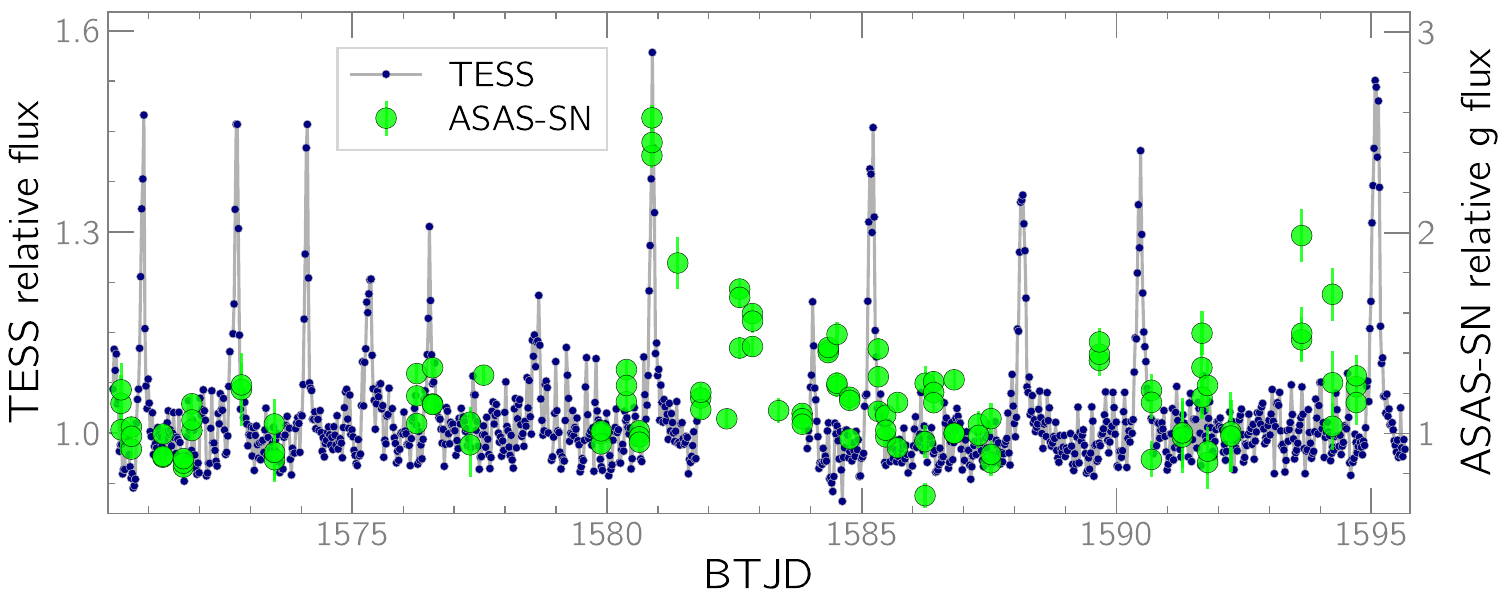}
    \includegraphics[angle=0,width=0.5\textwidth]{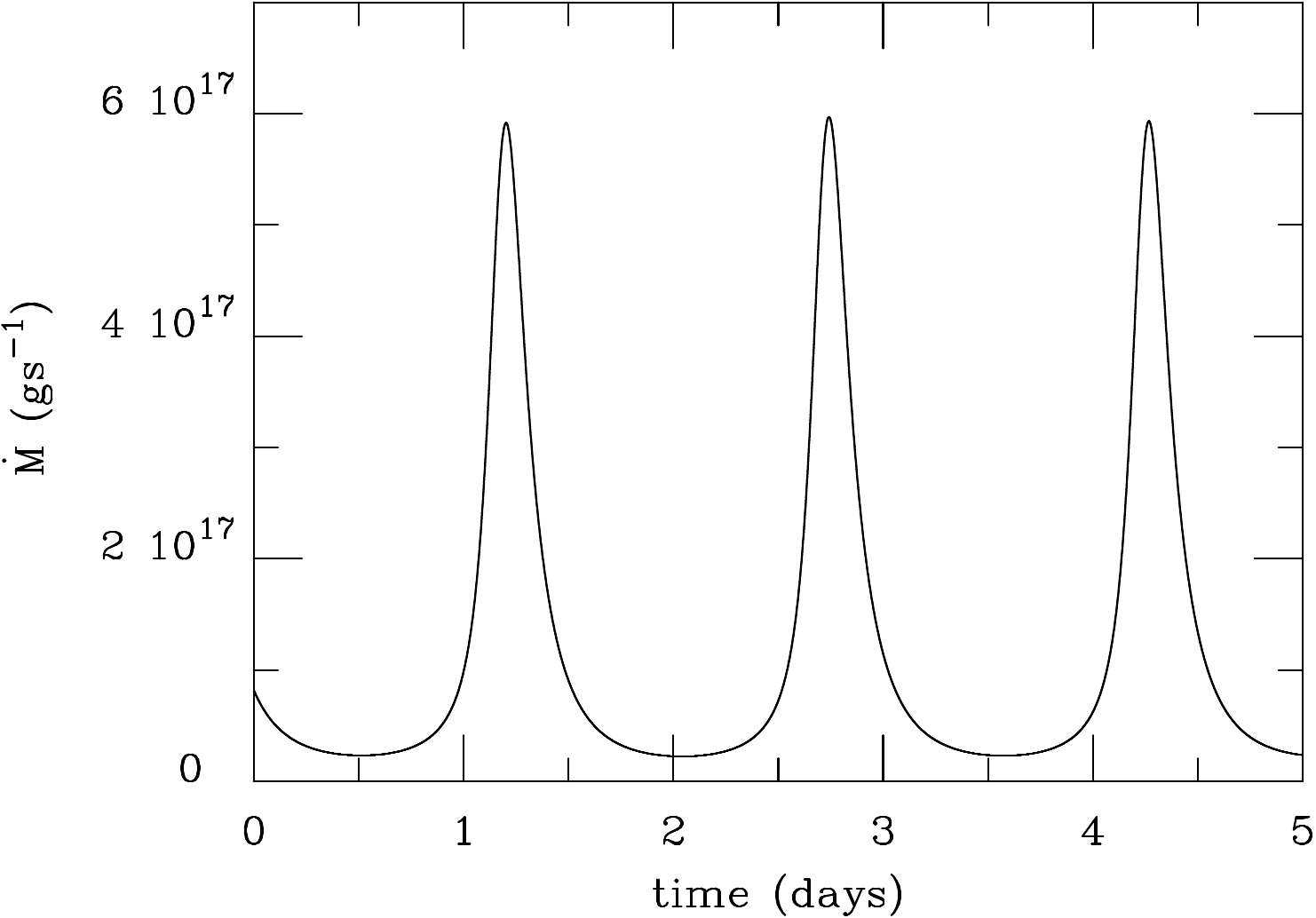}
    \caption{Top: TESS light curve of V1025 Cen, with simultaneous ASAS-SN observations overlaid on the secondary $y$-axis. Each TESS datum is a 30~min integration, while the ASAS-SN observations are 90~s exposures. Bottom: Example of a magnetic--gating instability light curve obtained with $\mu_{33} = 15$, $\Delta R/R_{\rm in} = 0.04$,
 $\Delta R_2/R_{\rm in} = 0.014$ and  $ \dot M_c = 1.24 \times 10^{17}\gms$ (see text).
({\sl Adapted from \citealt{Littlefield0122} and \citealt{Hameury0822}}).}
    \label{fig:1223}
\end{figure}

\citet{Scaringi1217,Scaringi0122} and \citet{Littlefield0122} proposed that the repetitive series of rapid, low-amplitude luminosity bursts in MV Lyr, TW Pic and V1025 Cen are produced by the magnetic--gating instability, originally proposed by \citet{Spruit0193} to explain the Type II X-ray bursts of the Rapid Burster.
This instability appears in systems with a magnetised accretor, when the inner disc radius, i.e. the radius of the magnetosphere, is close to the corotation radius $R_{\mathrm{cor}}$, defined as the radius at which the centrifugal forces on matter corotating with the white dwarf balance gravity forces:
\begin{equation}
    R_{\mathrm{cor}}=\left(\frac{GM}{{\Omega}^2_{\mathrm{spin}}} \right)^{1/3}   =3.52 \times 10^{10} M_1^{1 / 3} P_{\mathrm{spin}(\mathrm{h})}^{2 / 3} \mathrm{~cm}.
\label{eq:Rcor} 
\end{equation}
In such a situation, the rapidly rotating magnetosphere prevents accretion and causes the accretion flow to pile up just outside the magnetospheric boundary.
Eventually, this material compresses the magnetosphere until it is able to couple to the magnetic field lines (opening a gate in the magnetic wall) and accrete. Once the
reservoir of matter outside the magnetosphere is depleted, the cycle repeats itself, giving rise to episodic bursts of accretion. The magnetic--gating instability model (MGIM) was developed for magnetic neutron stars by \citet{Dangelo0810,Dangelo0911,Dangelo0212} (hereafter D'AS). However, there is a problem with such an interpretation of white-dwarf system outbursts: MV Lyr and TW Pic are not known to
be magnetic (although MV Lyr is a VY Scl star) and the presence a disc in V1025 Cen is uncertain \citep{Hellier0622}, while the MGIM assumes its existence.

\citet{Hameury0822} adapted the MGIM to the case of accreting magnetic white dwarfs and applied it to the {\sl bona fide} disc--possessing IP V1223.
The main uncertainty in the description of the disc-magnetosphere border is its width, the size of the region where the two interact. The most popular model, and still widely
used, is that of \citet{Ghosh1077,Ghosh0879,Ghosh1179} which assumes that the stellar fields invade the disc over a large range of radii.
However, the problem with this picture is that to make it work, one needs a very large (and unrealistic) magnetic diffusivity (see e.g. \citealt{Lai0114} and references therein).

Taking this into account, in their model D'AS correctly assume that the width of the disc-magnetosphere interaction region is narrow. 
They define a critical accretion rate
\begin{equation}
    \dot M_{c} = \frac{\varepsilon \mu^2}{4\Omega_{\rm spin}R^5_{\mathrm{cor}}} =2.63 \times 10^{14} P_{\mathrm{spin} {(\mathrm{h})}}^{-7 / 3}\, m^{-5 / 3} \mu_{33}^2 \gms,
\end{equation}
where $\varepsilon$ is a numerical factor describing the distortion of the
magnetic field by the disc. Following D'AS \citet{Hameury0822} took it to be equal to 0.1 (they call it $\eta$).  $\dot M_{c}$ is the rate
at which the inner (magnetospheric) disc radius is equal to the corotation radius. When the inner disc
radius is less than $R_{\mathrm{cor}}$, disc accretion proceeds in a standard way. In the opposite case, the accretion
rate is vanishingly small. The model depends on two parameters: $\Delta R$ -- the width of the disc--magnetospheric interaction,
and $\Delta R_2$ - the characteristic width of the change of accretion--rate through the interaction zone.
\citet{Hameury0822} adapted the D'AS model to the parameters of an IP, but in contrast to D'AS they used a standard definition of the viscosity
parameter $\alpha$ and applied it to configurations of thermally stable--hot and stable--cold discs. They found magnetic--gating instabilities
in both cases, but in the cold case the outburst amplitudes are too low to be of interest and the recurrence times are too long:
of the order of years.
\begin{figure}[ht]
    \centering
    \includegraphics[width=0.4\textwidth]{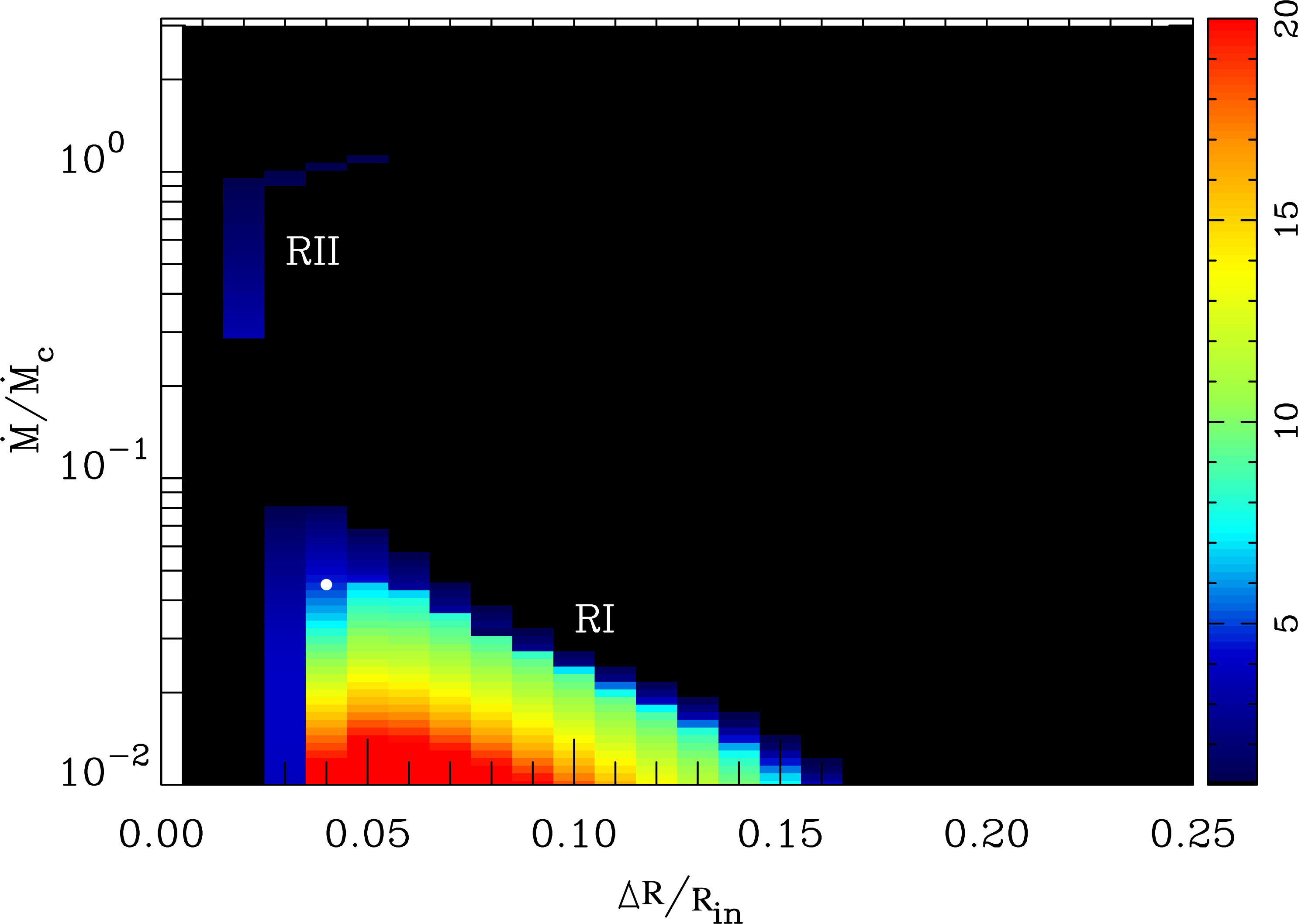}
    \includegraphics[width=0.4\textwidth]{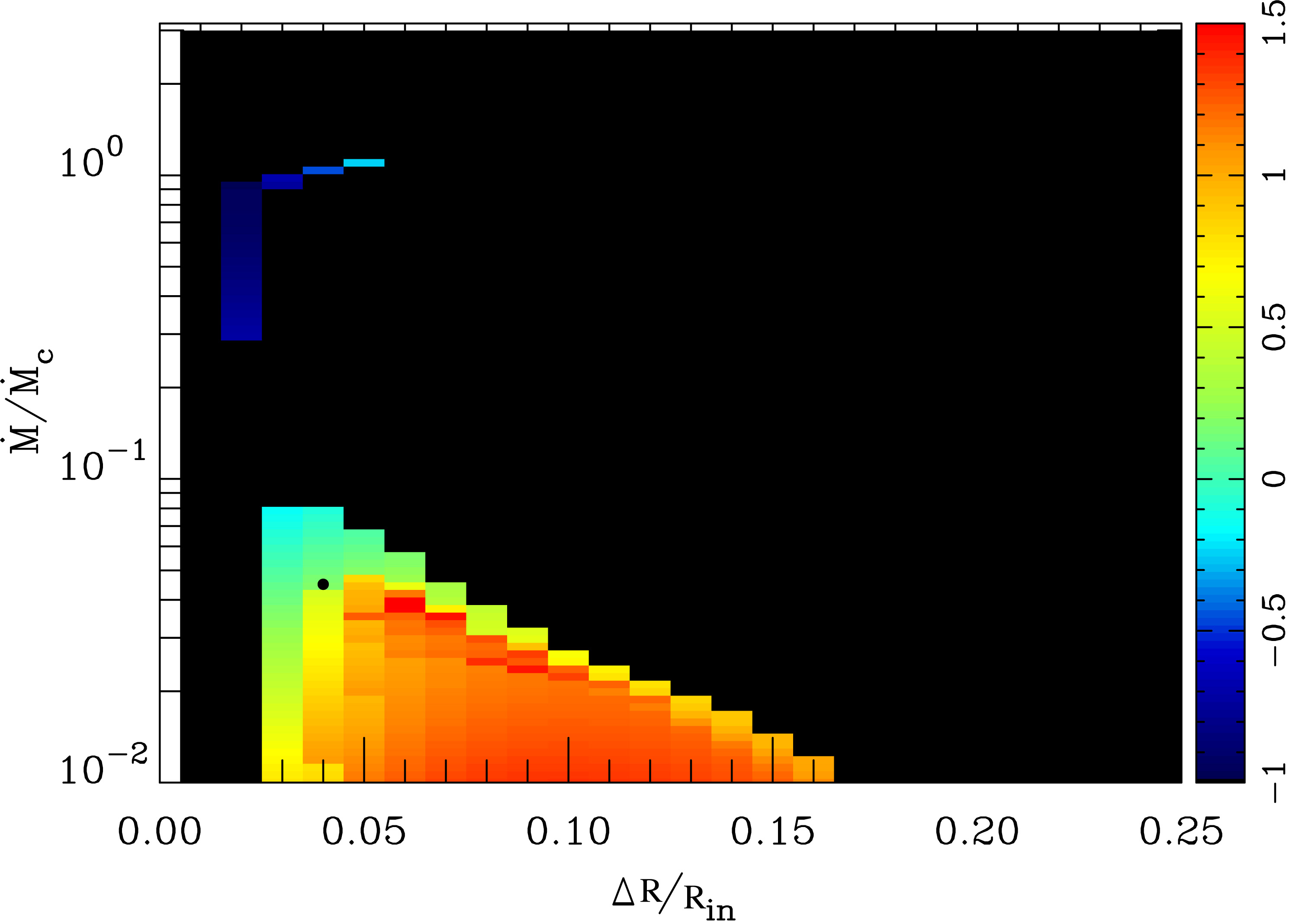}
    \caption{Instability map in the hot disc case for $ \mu_{33}=15$ and $P_{\rm spin}=0.2$ hr. Left: the colour scale shows the outburst amplitude defined as $A = \dot M_{\rm max}/\dot M_c$, with $\dot M_c = 2.8 \times 10^{18} \gms$. Right: the colour scale shows the logarithm of the recurrence time measured in days. Regions labelled RI and RII correspond to two regimes of the magnetic gating instability (see text). The white (left panel) and black (right panel) dots indicate the position corresponding to the light curve shown in Fig. \ref{fig:1223}. ({\sl From \citealt{Hameury0822}}).}
    \label{fig:gatestab}
\end{figure}

In the hot disc case, for mass--transfer rates $\dot M^+_{\rm crit} <\dot{M}_{\mathrm{tr}}< \dot M_c$, \citet{Hameury0822} find (as do D'AS) two regimes of magnetic--gating : RII and RI (see Fig. \ref{fig:gatestab}). In the RII region the outbursts have low amplitudes and short recurrence times. In this regime, the accretion rate varies smoothly with a period of a few hours, by less than a factor of  two. Regime RI, on the other hand, corresponds well to the sequence of short outbursts observed in V1223 Sgr. Fig.~\ref{fig:1223} (bottom) shows the variations of the accretion rate when  $\dot M_{\rm tr} = 0.045  \dot M_c = 1.24 \times 10^{17} \gms$ and $\Delta R/R_{\rm in} = 0.04$. These parameters correspond to the (white on the left, black on the right) point in the instability map in Fig. \ref{fig:gatestab}. The magnetic moment $\mu_{33}=15$ is more
appropriate for polars than for IPs, but was chosen in order to explore a large range of mass-transfer rates, reaching down to $0.01 \dot M_{c}$. For lower accretion rates (depending on $\Delta R$), the disc would become thermally unstable. \citet{Hameury0822} verified that simulations for $\mu_{33}=7$ produce a lightcurve similar to that of Fig. \ref{fig:1223}.
The MGIM  explicitly assumes the presence of an accretion disc. The similarities between the lightcurves of V1223 Sgr and the (possibly) discless IP V1025 Cen \citep{Littlefield0122} suggest that this model should also apply to systems with an accretion annulus (torus; see, e.g. \citealt{King1199}), instead of a disc.

It seems, however, that isolated, short outbursts that cannot be of the dwarf--nova type are not due to the MGIM. Although this instability can produce long recurrence times, of the
order of a month or more, the outburst duration would then also have to be long, because the ratio between the outburst recurrence time and its duration is roughly equal to the ratio between the mean mass-transfer and the peak--accretion rates, which is less than 30 in the model. We should also notice that the profiles of isolated outbursts often have a sharp rise and a slower decay, in contrast with the almost symmetric profiles produced by the MGIM and observed in V1223 Sgr and V1025 Cen. In fact, the shape of some isolated short bursts is similar to that of X--ray bursts from neutron stars. Hence the idea that they could be of thermonuclear origin.

\subsection{Micronov{\ae}}
\label{sec:McN}

\begin{figure}[ht]
\begin{center}
\includegraphics[width=8.0cm, height=6cm,angle=0]{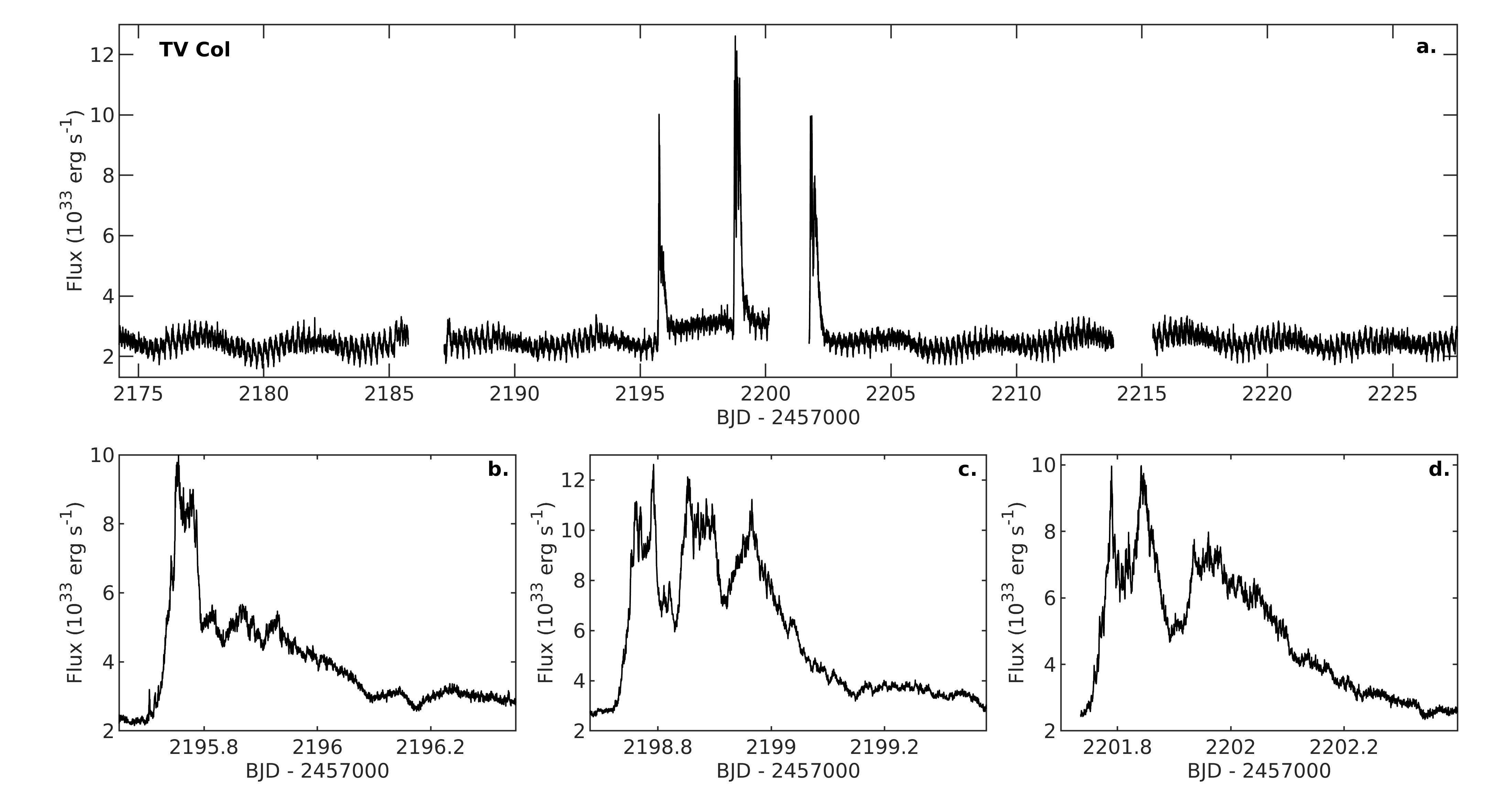}
\caption{Optical brightness variations in TV Col a. TESS lightcurve (20-s cadence) of TV Col. Panels b., c. and d. show 16.8 hours of data around the three detected rapid
bursts. The individual bursts yield integrated energies of $0.9 \times 10^{38}$erg, $1.6 \times 10^{38}$erg and $1.0 \times 10^{38}$erg, respectively. ({\sl From \citealt{Scaringi0422}})}
\label{fig:mnova}
\end{center}
\end{figure}

In several cataclysmic variable, TESS has observed short-duration fast-rise exponential-decay bursts, grouped in 
pairs or triples and lasting a few hours, with recurrence times of days to months. Such events are therefore different (mainly in recurrence--time properties) from the outbursts discussed in the previous section. \citet{Scaringi0422} noted such events in the IPs TV Col (see Fig. \ref{fig:mnova}), EI UMa and in the CV ASASSN-19bh. \citet{Schaeffer0522} discovered similar bursts in the recurrent nova V2487 Oph during quiescence. 

During these bursts the optical/UV luminosity
increases by a factor of more than $ 3$ in less than an hour and decays
during $\approx 10$ hours. Fast outflows with velocities  larger than $ 3500 \kms$, comparable
to the escape velocity from the white dwarf surface, have been observed in UV
spectral lines. The bursts have a total energy $\approx 10^{-6}$
that of classical nova explosions (``micronov{\ae}''), and their lightcurves are similar to those observed in Type I X-ray bursts. 
Guided by energy considerations and these similarities, \citet{Scaringi0422,Scaringi0722} proposed that these events result from thermonuclear runaway events in
magnetically-confined accretion columns on the surface of accreting white dwarfs. Thus the series of short outbursts observed in IPs would be the analogue of Type II X-ray bursts,
while those with longer recurrence times would be the equivalent of Type I X-ray bursts.

The model assumes the accretion column on the magnetic
poles of the white dwarf to be magnetically confined and 
increasing in mass. The pressure exerted on
the white--dwarf surface, resulting from the column's weight, causes its base
to sink to greater depths. If this magnetic confinement can hold until
the pressure at the base of the accreted column reaches $P_{\rm crit} \approx 10^{18}
\rm dyn\,/cm^2$, a thermonuclear runaway (TNR) will start and burn through most of the overlaying
accumulated mass in the column. The process can repeat every time
the pressure at the column's base reaches the critical pressure required to
drive a TNR.

For the flow to remain confined in the column by the magnetic pressure $P_{B} = {B^2}/{8\pi}$, the condition
\begin{equation}\label{eq:magConf}
\beta = \frac{P_{\rm base}}{P_{B}}<\beta_{\rm crit},
\end{equation}
must be satisfied, where
\begin{equation}
\beta = \frac{P_{\rm gas}}{P_{B}},    
\end{equation}
$P_{\rm gas}$ is the gas pressure of the magnetically confined material, and $\beta_{\rm crit}$ a critical value. 

As the weight of the column grows with time, the pressure at the base of the magnetically confined column ($P_{\rm base}$) will also grow, acting sideways on the magnetically confined boundary. When $\beta>\beta_{\rm crit}$ the column pressure substantially distorts the magnetic field lines, and plasma from the accretion column may spread on to the white--dwarf surface. 

The pressure exerted at the base of the accretion column is given by 
\begin{equation}\label{eq:Pbase}
    P_{\rm base}(t) =  \frac{{GM_{\rm WD}M_{\rm col}(t)}}{4\pi f R_{\rm WD}^4},
\end{equation}
where $M_{\rm WD}$ is the white--dwarf mass, $M_{\rm col}(t)=\dot{M}_{\rm acc} t$ the column mass, and
\begin{equation}
f=\left( \frac{R_{\rm col}}{2R_{\rm WD}} \right) ^2,
\end{equation}
where $R_{\rm WD}$ is the white--dwarf radius and $R_{\rm col}$ the radius of the circular accretion--column's footprint.
Then the accretion column will remain confined by the magnetic pressure $P_B$ as long as 
\begin{equation}\label{eq:magConf2}
\beta(t) = \frac{P_{\rm base}(t)}{P_B} < \beta_{\rm crit}
\end{equation}
where 
\begin{equation}
    \beta_{\rm crit} \approx 7\alpha^2
\end{equation}
and
\begin{equation}
    \alpha=\frac {R_{\rm col}} {h},
\end{equation}
with $h$ being the height of the accumulated material in the column \citep{Hameury0485}.

Assuming that outbursts are produced by the
freshly accreted hydrogen from the companion star, the CNO
cycle flash will yield $\sim 10^{16}$\,erg/g. Since the observed micronov{\ae} release 
$10^{38}$ -- $10^{39}$ erg, this implies column masses $5\times 10^{-12}\Msun \lesssim M_{\rm col} \lesssim 5\times 10^{-11}\Msun$, which for
a typical value $M_{\rm WD} \approx 0.8\Msun$
corresponds to a fractional accretion area $f \sim 10^{-6}$. This is a much smaller value than that deduced from observations, which suggest
rather $f \sim 10^{-2}$ -- $10^{-3}$ (see, \citealt{FKR2002} who, in a different context and for strongly magnetized polar systems\footnote{Accretion rates in polars might be too low to produce micronov{\ae}, on the other hand if such outbursts occur, they could have been missed.}, consider buried columns with $f \sim 10^{-7}$,
but find such a model ``contrived''.). For masses $M_{\rm WD}\gtrsim 1.3 \Msun$, $f$ can be increased to $10^{-4}$, reaching $10^{-3}$ for masses close to the Chandrasekhar limit.
Such high masses and strong fields  contradict the observationally determined white--dwarf's mass in TV Col: $0.74\Msun$.

The exploding--column model suffers  from two other drawbacks. First, it assumes that the magnetic field lines are solidly anchored in the white dwarf, at the
bottom of the accretion column. While such an assumption is justified in the case of a neutron star, it is rather uncertain when the accreting body is a white dwarf.
Second, plasma confined by a magnetic field is subject to instabilities that may lead to leakage preventing accumulation of the mass required to ignite a TNR. 

But what is the alternative? A model in which TNRs are triggered by magnetically confined ``blobs'', whose ram pressure reaches $P_{\rm crit} \approx 10^{18}
\rm dyn\,/cm^2$, requires even smaller fractional areas $f\sim 10^{-10}$ \citep{Scaringi0422}. Dwarf--nova outbursts are excluded, as are mass--transfer variations 
or reconnection events \citep{Scaringi0422}. In the end, the micronova explanation might be the best option.

\section{X-ray binaries}
\label{sec:XRB}

\subsection{X-ray transients}
\label{sec:XRT}

\begin{figure}[ht]
%\begin{wrapfigure}{r}[width=0.5\textwidth]
\begin{center}
\includegraphics[width=0.8\textwidth,height=7cm,angle=0]{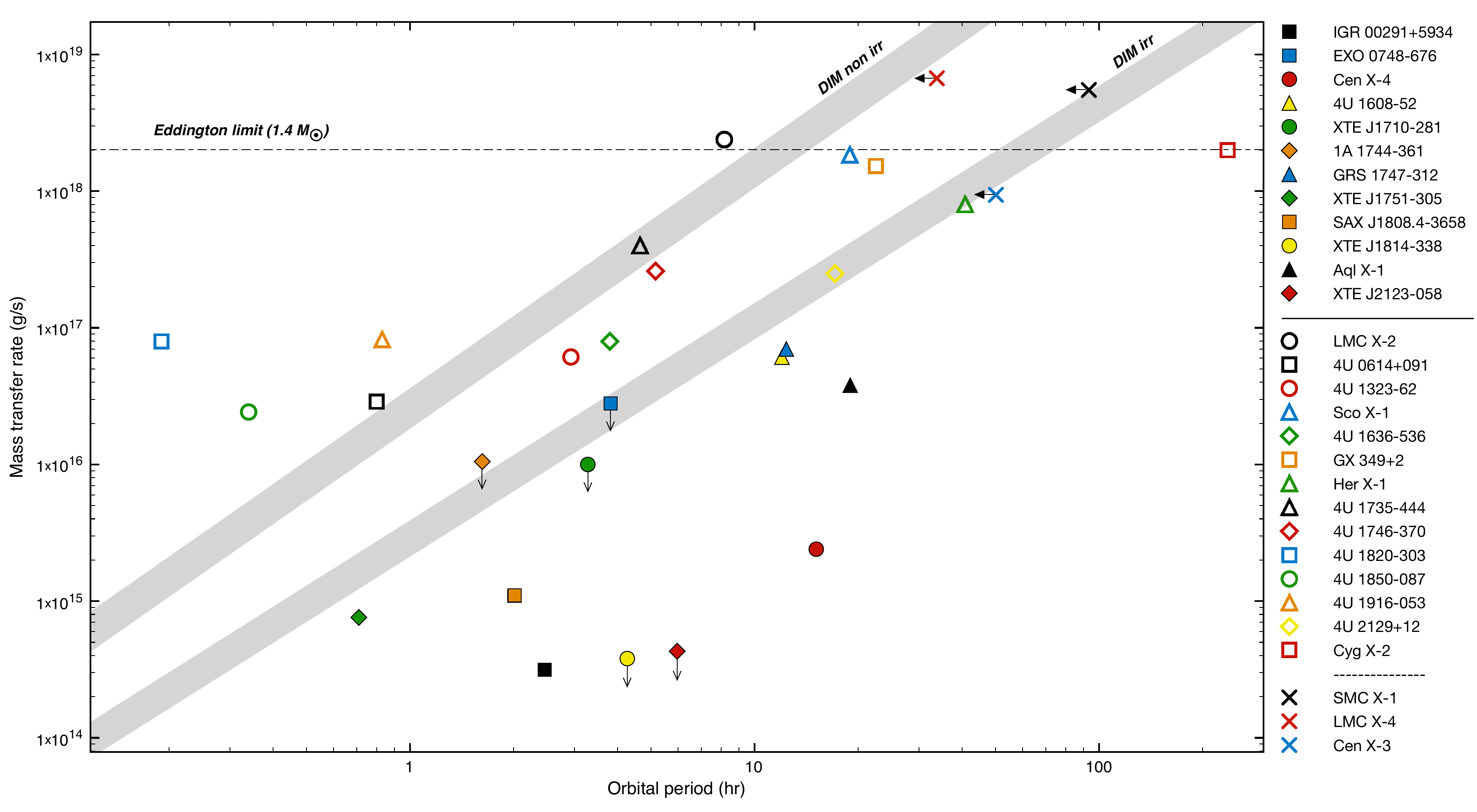}
\caption{
Mass-transfer rate as a function of the orbital period for X-ray binaries with neutron stars.  Transient and persistent LMXBs are indicated with filled and open symbols respectively, while the crosses indicate the high-mass, persistent systems. The shaded grey areas indicated by DIM irr and DIM non-irr represent the separation between the persistent (above) and transient systems (below) according to the DIM, respectively for irradiated and non--irradiated discs. The horizontal dashed line indicates the Eddington accretion rate for a $1.4 \Msun$ neutron star. The mass-transfer rate is assumed to be equal to the rate at
which the mass accumulates in the disc. {\sl (From \citealt{Coriat1208}}).}
\label{fig:NXRBstability}
\end{center}
\end{figure}

In X-ray binaries, the outer parts of the accretion disc are strongly X-ray irradiated by the central source \citep{vdP1094}. This heating effect must be taken into account when considering the disc thermal stability \citep{vdP0696}. In an irradiated disc the surface temperature is 
\begin{equation}
T_{\rm surf}^4 = \Teff^4 + T_{\rm irr}^4,
\end{equation}
with
\begin{equation}
 T_{\rm irr}  = \tilde{\mathcal{C}}\frac{\dot Mc^2}{4\pi R^2},
 \label{eq:cirr}
\end{equation}
where the irradiation constant  $\tilde{\mathcal{C}}$ \citep{Dubus1219} is usually taken to be $\sim 10^{-3}$ (\citealt{Dubus0107}).
Of course $\tilde{\mathcal{C}}$, as defined by Eq. (\ref{eq:cirr}), is unlikely to correspond to disc X--ray irradiation in all circumstances and this equation will have to be modified
when more is known about the process it is supposed to be describing (see \citealt{Tetarenko0218,Tetarenko0720}). In practice, even small modifications appear to be useful in some cases (see, e.g. \citealt{Coban0423}).

For an X-ray irradiated disc the stability criterion reads
\begin{equation}
\dot M > \dot{M}_{\text {crit }}=  9.5 \times 10^{14} \tilde{\mathcal{C}}_{-3}^{-0.36}
  m^{-0.64+0.08 \log \tilde{\mathcal{C}}_{-3}}  R_{10}^{2.39-0.10 \log \tilde{\mathcal{C}}_{-3}} \mathrm{~g} \mathrm{~s}^{-1},
\end{equation}
where $\tilde{\mathcal{C}}=10^{-3}\tilde{\mathcal{C}}_{-3}$, $R= R_{10}10^{10}\rm cm $ and a factor very weakly depending on the viscosity parameter $\alpha$ has been dropped.

Figure \ref{fig:NXRBstability} \citep{Coriat1208} shows the stability properties of neutron--star X--ray binaries. The non--irradiated--disc criterion is also plotted. Clearly, the irradiated--disc stability limit correctly separates steady systems from outbursting ones, i.e. X-ray transient systems. 

While  both steady and transient systems are found among low--mass, neutron--star X--ray binaries, all black--hole, low--mass X-ray binaries are transient. The reason for this difference is  still unknown \citep{King0696,Justham0306,Yungelson0806,Yungelson0908,Wiktorowicz0914}, but  all such black--hole X--transients also lie below the irradiated--disc stability criterion \citep{Coriat1208}.

\subsection{Lightcurves of X-ray transients}

As for dwarf nov{\ae}, and for the same reason, in X-ray transients systems the simplest version of the DIM predicts fast--rise and slow--decay outbursts, but contrary to the common opinion \citep[e.g.][]{King0198,Dubus0107,Lasota0315} the decay parts of the outburst  lightcurves produced by the irradiated--disc instability are {\sl not} exponential.
As shown by \citet[][see also \citealt{King0698} and \citealt{Lipunova0400}]{Ritter0101}, according to the model, the initial (irradiation controlled) decay from outburst maximum of an X--ray transient  is described by
\begin{equation}
\dot{M}=\dot{M}_{\rm max} \left[ 1+ \frac{t}{t_0} \right]^{-10/3},  
\label{eq:varmdot}
\end{equation}
where $\dot{M}_{\rm max}$ is the accretion rate at the outburst maximum and $t_0$ is given by: 
\begin{equation}
\label{eq:tzero}
    t_0 = 3.19 \,\alpha_{\rm 0.2}^{-4/5} M_1^{1/4} R_{12}^{5/4} \dot{M}_{\rm max,19}^{-3/10} \; \rm yr,
\end{equation}
where $R=R_{12}10^{12}\cm $, $\dot{M}_{\rm max,19} = \dot{M}_{\rm max}/10^{19}$~g\,s$^{-1}$ \citep{Hameury1120}.
\begin{figure}[ht]
    \centering
    \includegraphics[trim= 0 2.5cm 0 0, clip, width=0.5\textwidth]{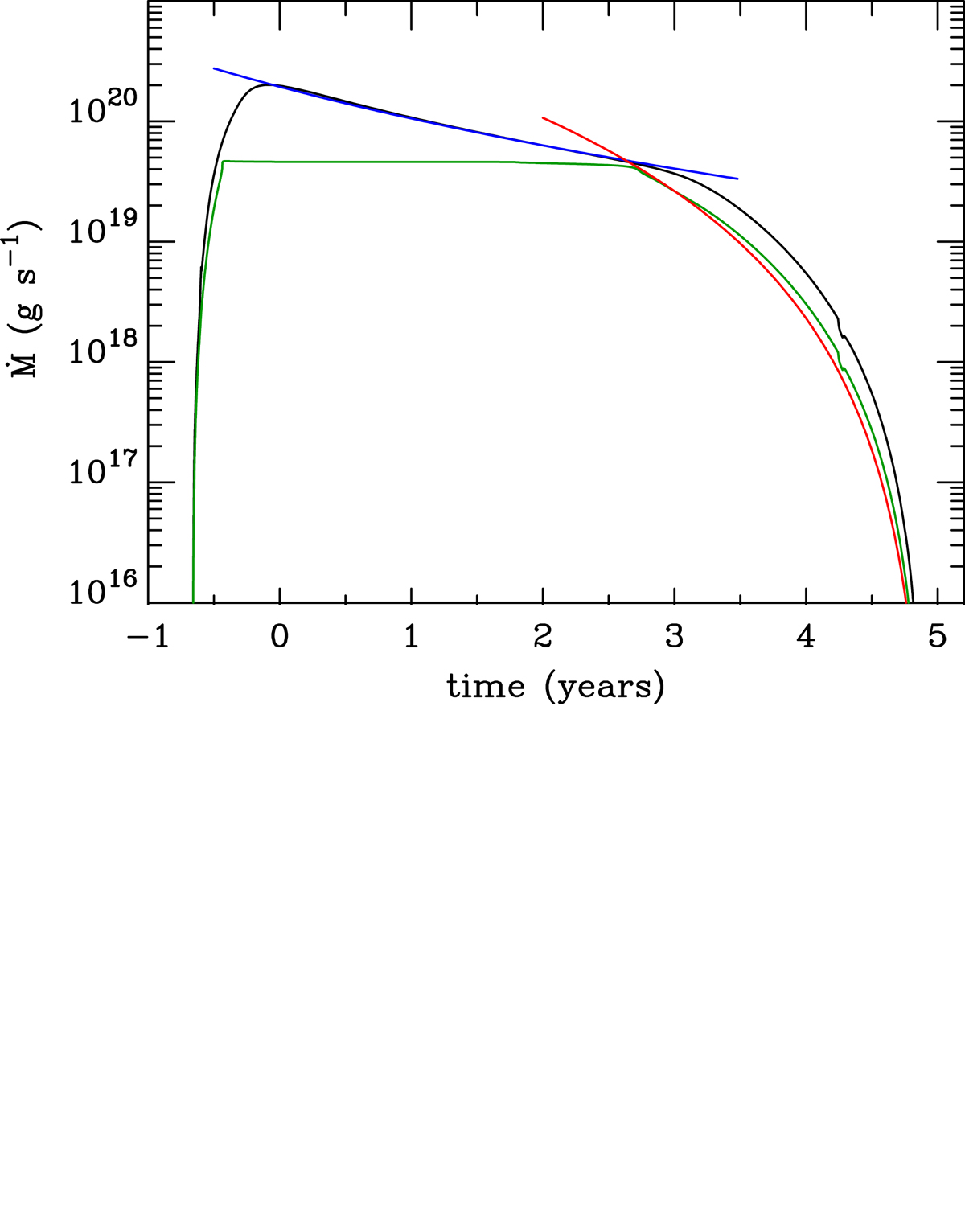}
    \caption{This figure shows the mass accretion rate calculated numerically (black curve) and the critical rate $\dot{M}_{\rm crit}^+(R_{\rm out})$ (green curve). The blue curve gives the analytic solutions (Eq. \ref{eq:varmdot}) when the entire disc is in the hot state, and the red curve corresponds to the outburst decay due to a propagating cooling front (Eq. \ref{eq:decay}). ({\sl Adapted from \citealt{Hameury1120}}).}
    \label{fig:lccompare}
\end{figure}
$t_0$ corresponds to the time it takes the accretion rate to fall to one tenth of its initial value.

Based on Eq. (\ref{eq:varmdot}) and the DIM, \citet{Hameury1120} found analytical formul{\ae} describing the decay lightcurves of X--transients which can be used to determine
the disc parameters.

The characteristic timescale of the disc evolution $\tau$ is defined as
\begin{equation}
    \tau = \frac{M_d}{\dot{M}_{\rm max}} = 0.92 f^{-0.3} M_1^{0.37} f_{\rm irr}^{0.15} R_{12}^{0.62} \alpha_{0.2}^{-0.8} \; \rm yr, 
\label{eq:tau}
\end{equation}
where\footnote{There is some confusion in the literature about the definition of the ``irradiation constant'' ${\mathcal{C}}$: on the one hand \citet{Dubus9902} define ${\mathcal{C}}$ (called $\tilde{\mathcal{C}}$ here) through Eq. (\ref{eq:cirr}) (this definition is also used by \citealt{Coriat1208}), on the other, \citet{Dubus0107}, ``extracting'' the accretion efficiency,  in Eq. (\ref{eq:cirr}) use $\eta{\mathcal{C}}$, instead of $\tilde{\mathcal{C}}$. Finally, \citet{Hameury0420} find it convenient to use $f_{\rm irr}$, as in Eq.(\ref{eq:cirr}). As the co--author of three of the above--mentioned papers, I would like to apologise for this inconvenience.}
\begin{equation}
\label{eq:firr}
  f_{\rm irr} = 0.2\, \tilde{\mathcal{C}} = \frac{\eta\mathcal{C}}{5\times 10^{-4}}
\end{equation}
and the ratio of the maximum to the critical accretion rate $f=\dot{M}_{\rm max}/\dot{M}_{\rm crit}^+(R_{\rm out}) > 1$ is given by
\begin{equation}
   f \sim \phi(\alpha_{\rm h}/\alpha_{\rm c}) \dot{M_{\rm tr}}/ \dot{M}_{\rm crit}^+(R_{\rm out}),
\end{equation}
where $\dot M_{\rm tr}$ is the mass transfer rate from the secondary, $\phi(\alpha_{\rm h}/\alpha_{\rm c})\equiv \dot M_{\rm max}/ \dot M_{\rm tr}$, while $\alpha_{\rm h}$ and $\alpha_{\rm c}$ are respectively the viscosity parameter in the hot and cold disc.
As in the case of dwarf nov{\ae}, for the outbursts to have the observed amplitudes, this ratio has to satisfy the condition $\alpha_{\rm h}/\alpha_{\rm c} \approx$ 4--10 \citep{Hameury0898,Dubus0107}. For $\alpha_{\rm h}/\alpha_{\rm c}=5$, $\phi \approx 25 - 35$; for $\alpha_{\rm h}/\alpha_{\rm c}=10$, $\phi \approx 70 - 100$.
Since the necessary condition for disc instability is $\dot{M_{\rm tr}}/ \dot{M}_{\rm crit}^+(R_{\rm out}) < 1$, it  follows that $f<\phi$.
The mass $M_d$ of a fully hot disc is obtained by integrating the critical surface density over the disc \citep{Hameury1120}:
\begin{equation}
    M_{\rm d} = 4.3 \times 10^{26} \alpha_{0.2}^{-4/5} \dot{M}_{19}^{7/10} M_1^{1/4} R_{12}^{5/4} \; \rm g.
    \label{eq:md}
\end{equation}

The critical accretion rate below which a hot, irradiated disc become unstable is given by
\begin{equation}
    \dot{M}_{\rm crit}^+ \approx 2.4 \times 10^{19} M_1^{-0.4}f_{\rm irr}^{-0.5} \left(\frac{R_{\rm fr,max}}{10^{12}\cm}\right)^{2.1} \; \rm g \, s^{-1},
\label{eq:mdotcrit}
\end{equation}
(\citealt{Lasota0106}; slightly different fits to the critical values of the disc parameters are found in \citealt{Lasota0808}).  

When the front fails to reach the outer disc edge\footnote{In X-ray binary transients outbursts are only of the inside-out type \citep{Lasota0106}.}, $\dot{M}_{\rm max}$, the maximum accretion rate during an outburst,  is roughly equal to the rate at the maximum distance reached by the transition front $R_{\rm fr,max}$ \citep{Lasota0315}, because at the outburst peak, the portion of the disc that has been brought into the hot state is almost steady, and the mass accretion rate is thus equal to the minimum (critical) rate $\dot{M}_{\rm crit}^+$ for which a hot stable disc can still exist. Only when the heating front reaches the outer disc edge ($R_{\rm fr,max} \approx R_{\rm out}$) is the ratio $\dot{M}_{\rm max}/\dot{M}_{\rm crit}^+(R_{\rm out}) > 1$. 

\begin{figure}[ht]
    \centering
    \includegraphics[width=0.8\textwidth]{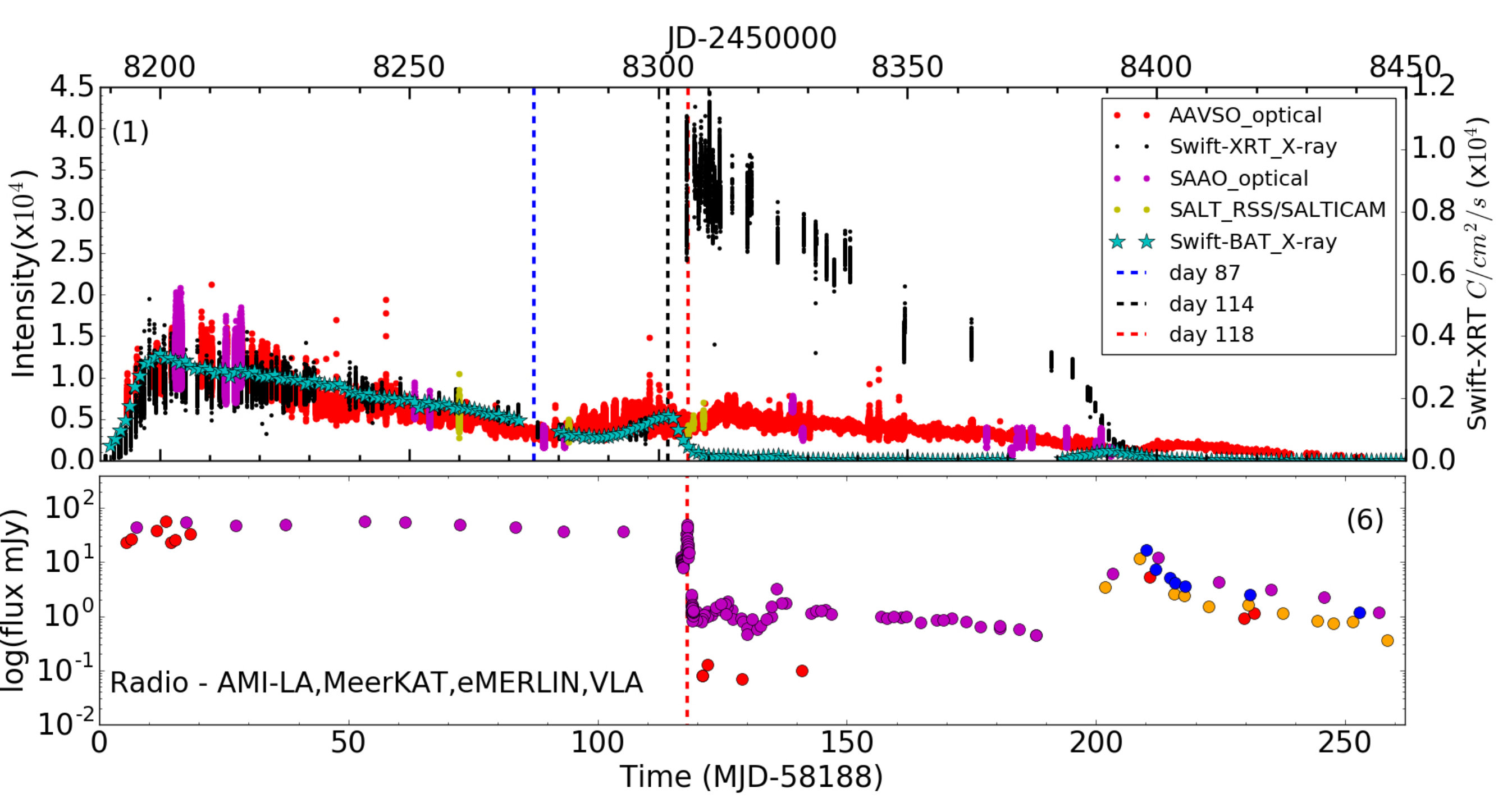}
    \caption{Top: Multiwavelength light curves of the black--hole X--ray transient MAXI J1820+070. From top to bottom: (1) Optical (AAVSO red; SAAO magenta; SALT yellow) and X-ray Swift XRT
black; Swift BAT cyan. Optical magnitudes were converted to an arbitrary intensity scale; Swift XRT count rates (right y-axis) are offset by $+9000$. 
Bottom: Radio fluxes
from AMI to LA (magenta; 15.5 GHz), MeerKAT (orange; 1.28 GHz), eMERLIN (red; 1.5, 5 GHz), and VLA (blue; C band) are plotted in the bottom panel
(see Bright et al. 2020 and Homan et al. 2020 for details). Three vertical dashed lines mark key moments during this outburst: Days 87 (blue), the beginning of
large amplitude optical modulations; 114 (black), peak of the Swift/BAT secondary maximum; 118 (red), time of the AMI radio flare, jet ejection and X-ray
state change from hard to soft. ({Adapted from \citet{Thomas0122}).}}
    \label{fig:1820}
\end{figure}

The maximum accretion rate can be related to the characteristic decay time by combining Eqs. (\ref{eq:mdotcrit}) and (\ref{eq:tau}):
\begin{equation}
    \dot{M}_{\rm max} = 2.0 \times 10^{19} \alpha_{0.2}^{2.71} f^{2.02} M_1^{-1.65} \left( \frac{ \tau}{1 \; \rm yr}\right)^{3.39} f_{\rm irr}^{-1.01}  \; \rm g \, s^{-1}.
\label{eq:mdot-tau}
\end{equation}
When both $\dot{M}_{\rm max}$ and $\tau$ are known  from observations, this relation determines $f$, and hence $\dot{M}_{\rm crit}^+$ and the size of the accretion disc. When this size can be estimated (from the orbital period and mass ratio) this can be used to estimate $\alpha_h$.

Alternatively, from Eqs. (\ref{eq:varmdot}), (\ref{eq:tzero}) and (\ref{eq:mdotcrit}) we can obtain the duration of the the quasi--steady phase of the outburst decay, during which 
$\dot{M}/\dot{M}_{\rm crit}^+(R_{\rm out}) > 1$;
\begin{equation}
    \Delta t_1=t_0\left[1.38 t_0^{-0.50} M_1^{0.25} \dot{M}_{19, \max }^{0.15} f_{\mathrm{irr}}^{0.15} \alpha_{0.2}^{-0.4}-1\right],
    \label{eq:deltat}
\end{equation}
from which we can determine $\alpha_h$ if the accretor mass is known, since here the dependence on $f_{\mathrm{irr}}$ is weak.

Once the decreasing accretion rate reaches the critical value, a cooling front starts propagating (always inwards, since in the hot state the accretion rate is roughly constant but the critical values increase with radius) and switches off the outburst. Since now the hot disc is shrinking we have
\begin{equation}
\dot{M}_{\rm d} = -\dot{M} - \dot{M}_{\rm fr}+ 2 \pi R_{\rm fr} \Sigma \dot{R}_{\rm fr}, 
\end{equation}
where $ R_{\rm fr}(t)$ is the front positional radius and $\dot{M}_{\rm fr}$ is the mass flow at the propagating--front position\footnote{{\it Not} the mass transfer rate.}.
Since $\dot{M}=\dot{M}_{\rm crit}^+(R_{\rm fr})$, from Eq. \ref{eq:mdotcrit}, the radius $R_{\rm fr}$ can be expressed as a function of $\dot{M}$:
\begin{equation}
\dot{M}_{\rm d} = -2.47 (\dot{M} + \dot{M}_{\rm fr}) =  \xi \dot{M}.  
\label{eq:mdot1}
\end{equation}
Putting $\xi=6.3$ \citep{Hameury1120}, from Eq. (\ref{eq:md}) using Eqs. (\ref{eq:mdotcrit}) and (\ref{eq:mdot1}); we obtain:
\begin{equation}
    \dot{M} = 6.7 \times 10^{19} \alpha_{0.2}^{2.71} M_1^{-1.65} f_{\rm irr}^{-1} \left[\frac{(t_{0}^{\prime}-t)}{1 \; \rm yr} \right]^{3.39} \; \rm g \, s^{-1},
    \label{eq:decay}
\end{equation}
where $t_0^{\prime}$ is a constant that is determined by the condition that, when the cooling front starts, $\dot{M}$ is equal to $\dot{M}_{\rm crit}^+$ at the maximum front--transition radius. $t_0^\prime$ can then be written as:
\begin{equation}
    t_0^{\prime}=0.7\, M_1^{0.37} f_{\rm irr}^{0.15} \alpha_{0.2}^{-0.8} r_{12}^{0.62} \; \rm yr. 
    \label{eq:t0}
\end{equation}
As can be seen in Fig. \ref{fig:lccompare}, Eq. \ref{eq:decay} represents the results of numerical simulations  quite well also for this part of the lightcurve.
The shift between the two curves is due to the fact that in the simulations the accretion rate is not exactly equal to the critical rate.

Using Eqs. (\ref{eq:mdot-tau}), (\ref{eq:deltat}), (\ref{eq:decay}) and (\ref{eq:t0}) we can determine the properties 
of the outbursting accretion disc from observed lightcurves, in particular the value $\alpha_h$\footnote{Such a  method was used by \citet{Tetarenko0218,Tetarenko0720} but in those papers it was assumed that 
the quasi--steady decay phase is exponential.}. In Sect.\ref{sec:TULX}, I will show the results of applying this method to transient ultraluminous X--ray sources.

As in the case of dwarf nov{\ae}, quite often the real lightcurves of X-ray transients (see Fig. \ref{fig:1820}) are much more complicated than those predicted by the DIM (Fig. \ref{fig:lccompare}). The \citet{Hameury1120} model assumes a flat disc, while the lightcurve in Fig. \ref{fig:1820} requires the disc to be warped \citep{Thomas0122}. Also, this model describes neither the source of the hard X--ray component(s) (the corona), nor the jet ejected at some phase of the outburst.
Figure \ref{fig:lccompare} compares the evolution of the accretion rate as found in numerical simulations with the analytical estimate given by Eq. (\ref{eq:varmdot}), demonstrating how well the analytical formula represents the results of numerical simulations.
One should stress again that, although the decay is not far from exponential, the ``$-10/3$'' power law is  a much better fit by far.

%Figure \ref{fig:Mdot-rtr} shows all our models in the plane $(r_{\rm tr,max},\dot{M}_{\rm max})$ where $\dot{M}_{\rm max}$ has been normalized by $M_1^{0.4} f_{\rm irr}^{0.5}$.  We have omitted models 4 and 5, in which irradiation is not taken into account, as well as model 24 in which $f_{\rm irr}$ varies.

\section{Ultraluminous X-ray sources}
\label{sec:ULX}

By definition, ultraluminous X-ray sources (ULXs) have luminosities $L_X > 10^{39}\ergs$ and are not located in galaxy centres. The defining luminosity has no
astrophysical basis but was chosen for corresponding roughly to the Eddington luminosity of a $10\msun$ black hole, this critical luminosity
being defined as
\begin{equation}
\label{eq:Ledd}
\Ledd = 1.3 \times 10^{38} \left(\frac{M}{\Msun}\right)\, {\rm erg\,s^{-1}},   
\end{equation}
while the corresponding Eddington accretion rate is given by
\begin{align}
\Mdoted \equiv \frac{\Ledd}{\eta c^2} & = 1.4\times 10^{18}\eta_{0.1}^{-1} \left(\frac{M}{\Msun}\right)\,{\rm g\,s^{-1}} \\
& = 2.2 \times 10^{-8}\, \eta_{0.1}^{-1} \left(\frac{M}{\Msun}\right) \rm M_{\odot}\,\rm yr^{-1},
\end{align}
where $\eta=0.1\eta_{0.1}$ is the radiative efficiency of accretion.
ULXs were identified as a separate class of objects at the end of the previous millennium \citep{Colbert9907}. After about fifteen years, during which the majority of researchers
involved in their study believed that they contained intermediate--mass black-holes (IMBH), the discovery by \citet{Bachetti1410} that the source ULX-2 in the galaxy M82 is a pulsar confirmed the view of the dissenters \citep{King0501}, who had claimed from the start  that ULXs are just a phase in the life of (presumably massive) X-ray binaries containing stellar--mass accreting objects, i.e. black holes or neutron stars, or even white dwarfs. 

Now we know that out of the $\sim$ 1800 observed ULXs (see \citealt{king0623} and references therein) at least 10 contain magnetized neutron stars, detected through their periodic pulses (PULXs; see Table \ref{tab:pulxs}). Four of them are transient: they are members of Be--X binary systems, which become X--ray sources when the eccentric orbit of the compact companion (in most cases a neutron star) of the massive Be star crosses its circumstellar disc. In most cases this disc--crossing produces sub-Eddington--luminosity outbursts (called ``Type I''), but from time to time, most probably due to von Zeipel-Kozai-Lidov oscillations \citep{Martin0914}, it results in a giant (super-Eddington; ``Type II'') outburst. 

Since the maximum mass of neutron star is $\sim 2 \Msun$, it is clear that the {\sl observed} luminosity of all PULXs is super-Eddington - up to 1000 times larger than the Eddington luminosity of $1\Msun$. Since at super-Eddington accretion rates the resulting luminosity is
\begin{equation}
\label{eq:superedl}
L = \Ledd\left[ 1 + \ln\frac{\dot M}{\Mdoted} \right],  
\end{equation}
(see e.g. \citealt{Lasota16,king0623}), it is clear that, for realistic values of the mass--transfer rate in stellar binary systems ($\lesssim 10^4\Mdoted$, say), observed luminosities in excess of $10\Ledd$ ($\sim 2\times 10^{39}\ergs$) cannot be intrinsic if the critical luminosity of the accretion flow is equal to the Eddington luminosity. Since strong magnetic fields lower the Thomson scattering opacity \citep{Canuto0571}, in their presence the critical luminosity (corresponding to the equality of the radiative and gravitational forces) is equal to
\begin{equation}
\label{eq:Lcrit}
{L_{\rm crit}} \approx 2 B_{12}^{4/3}\left(\frac{g}{2\times 10^{14}\rm cm\,s^{-2}}\right)^{-1/3}{\Ledd},
\end{equation}
(when $L_{\rm crit} \gg \Ledd$ ) where $g=GM/R^2$ , as shown by \citet{Paczynski0792}.
Therefore, to be intrinsic, the observed PULX luminosities $\gtrsim 10^{40}\ergs$ must be emitted by a plasma permeated by magnetar--strength fields
$> 10^{14}\rm G$. As an alternative, not requiring super--strong magnetic fields, it has been proposed that a buoyancy-driven, ``photon bubble'' wave pattern can facilitate the
escape of radiation, producing intrinsic super-Eddington luminosities from accretion columns (\citealt{Begelman1017}, see also \citealt{Arons0492}). However, the
reality of this process has yet to be confirmed and it seems that very super--Eddington emission from such a bubbling column would also require beaming.

In the absence of super--strong (magnetar--strength)  magnetic fields, the emitted luminosity {\sl must be} beamed and it is the apparent luminosity that is super-Eddington,
Eq. (\ref{eq:superedl}) becoming
\begin{equation}
\label{eq:superedlbeam}
L_{\rm app} = \frac{1}{b}\Ledd\left[ 1 + \ln\frac{\dot M}{\Mdoted} \right],  
\end{equation}
where the beaming factor $b < 1$, as  already proposed by \citet{King0501}.
There are thus three options explaining the observed hugely super-Eddington luminosities of PULXs:
\begin{itemize}
    \item Magnetar--strength fields of accreting neutron stars (\citealt{Dalosso0515,Eksi0315,Bachetti1022}).
    \item Buoyancy--driven, ``photon bubble'' wave patterns allowing the escape of radiation at extremely super-Eddington rates and some amount of geometrical beaming \citep{Begelman1017}.
    \item Geometrical beaming (collimation) by an accretion--flow wind \citep{King1605,King1702,King0519,King0520,king0623}.
\end{itemize}

I will show, however, that the magnetar hypothesis can be easily rejected \citep{Lasota1223}: a fundamental property of PULX, the value of their spin-up rate $\dot \nu$ ($\nu$ is the pulsar's spin frequency), makes the first option physically impossible, as first pointed out by \citet{Kluzniak1503}. The photon--bubble mechanism which allows large super--Eddington intrinsic luminosities has not been sufficiently developed to be compared to observations, but would require some beaming anyway \citep{Begelman1017}.

\begin{figure}
\begin{center}
\includegraphics[width=12.0cm,height=6.5cm,angle=0]{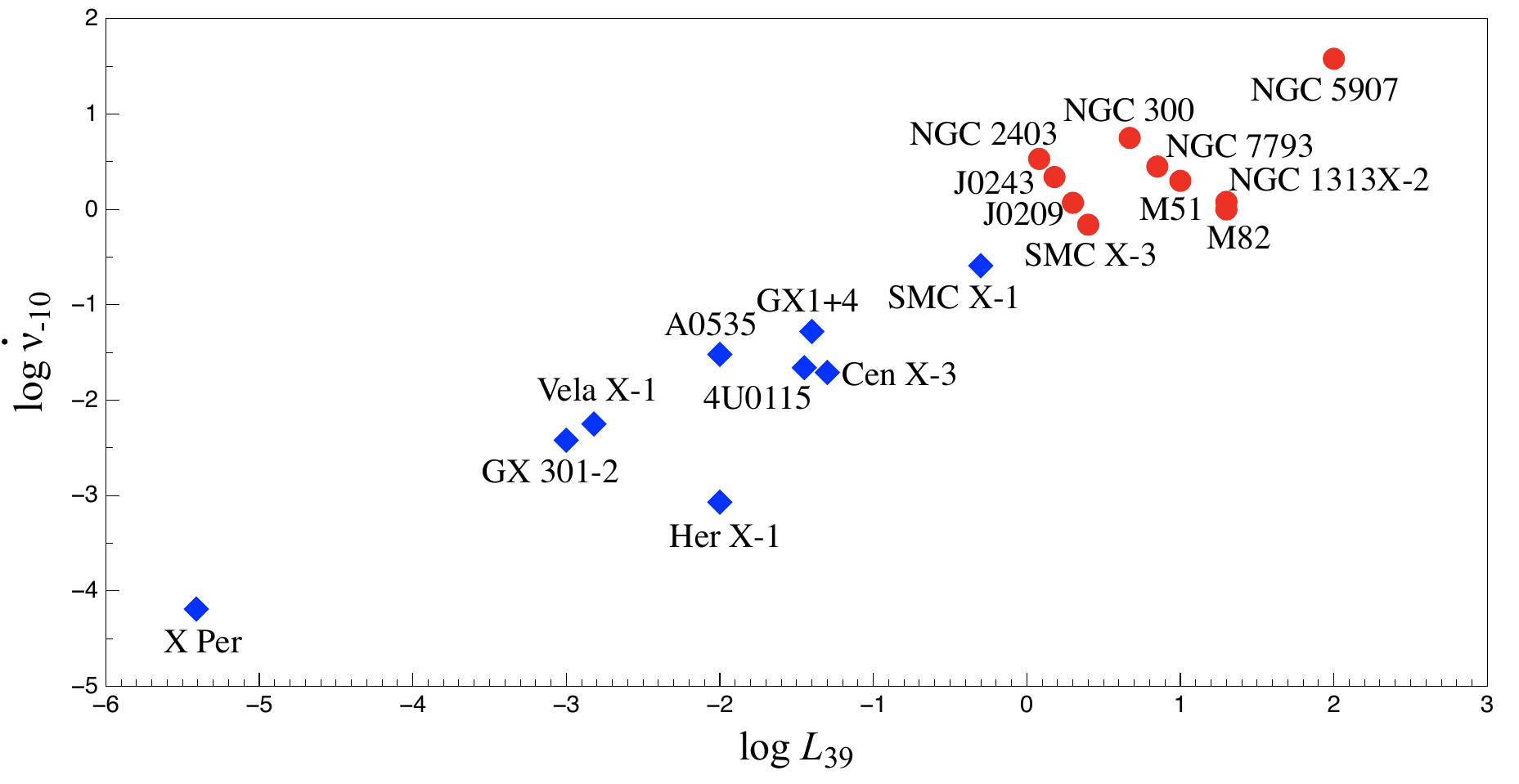}
\caption{The $L_{39}$ -- $\dot \nu_{-10}$ diagram for XRPs and PULXs. Red dots: the ten PULXs with known spin-up rates. Blue diamonds:
selected (for comparison) sub--Eddington--luminosity X-ray pulsars. ({\sl For details see \citealt{king0623}})}
\label{fig:lnupulx}
\end{center}
\end{figure}

Figure \ref{fig:lnupulx} shows that the spin-up rate $\dot \nu$ for {\sl all} (with sub-- and super--Eddington luminosities) X-ray pulsars is strongly correlated
with their X--ray luminosity $L_X$. This correlation (over-seven-orders of magnitude in luminosity)  can be explained as resulting from the domination of the accretion torque over all other torques acting in these systems:
\begin{equation}
\dot\nu = \frac{\dot J(R_M)}{2\pi I} = \frac{\dot M (GMR_M)^{1/2}}{2\pi I} \propto \dot M^{6/7}\mu^{2/7},
\label{eq:dotnudef}
\end{equation}
where $R_{\rm M} \propto \dot M^{-2/7}\mu^{4/7}$ (Eq. \ref{eq:rm})  is the magnetospheric radius, $\mu$ the neutron star's magnetic moment and $I$ the neutron star's moment of inertia.
Figure \ref{fig:lnupulx} shows that PULXs are characterised not only by their luminosity ($L_X > 10^{39}\ergs$) 
but also by their spin-up rate ($\dot \nu \gtrsim 10^{-10} \rm s^{-2}$): they are all located inside a rectangle delimited by the
values of these two quantities. If we did not know the distances to the X-ray pulsars, the value of their spin-up rate alone 
would allow us to separate ``normal'' XRPs from PULXs. Indeed, since the transient ULX~Swift~J0243.6+6124
is in the Galaxy (the only known such source there), its exact distance is uncertain, but while its deduced luminosity is close to the
ULX-defining limit ($\sim 1.5 \times 10^{39} \ergs$), the value of its spin-up rate ($2.2 \times 10^{-10} \rm s^{-2}$) puts it safely into the PULX category.

The magnetospheric radius is defined by the equation \citep{FKR2002}
\begin{equation}
R_{\rm M} = 2.6 \times 10^8 q\, \left(\frac{\dot M}{10^{17}\gms}\right)^{-2/7} \left(\frac{M}{\Msun}\right)^{-3/7} \mu_{30}^{4/7}\, \rm cm,
\label{eq:rm}
\end{equation}
where $q\sim 1$ is a factor taking into account the geometry of the accretion flow at the magnetosphere and $\mu=10^{30} \mu_{30}\rm Gcm^3$. 
Putting $M\approx 1\Msun$ and $q\approx 1$, from Eqs. \ref{eq:rm} and \ref{eq:dotnudef} one obtains 
\begin{equation}
\label{eq:mdotnudot}
    \dot M \approx 5.7 \times 10^{18} \dot\nu^{7/6}_{-10} \mu_{30}^{-1/3} \gms.
\end{equation}
In general, super--Eddington luminosities are not proportional to the accretion rate.
Only in the presence of very strong magnetic fields, when $L_{\rm crit} \gg \Ledd$ (see Eq. \ref{eq:Lcrit}) can we assume that $L_X \approx 0.1 \dot M c^2$,
since $\Ledd$ is no longer the critical luminosity.
This would be the case if PULXs contained accreting magnetars.

In such a {\sl sub--critical} case, from Eq. (\ref{eq:mdotnudot}), one gets  the relation
\begin{equation}
\label{eq:Lxnu}
 L_X \approx 2 \times 10^{38} \dot\nu^{7/6}_{-10} \mu_{31}^{-1/3} \ergs \approx \Ledd.
\end{equation}
But this contradicts the condition $L \gtrsim L_{\rm crit}\gg\Ledd$, assumed in its derivation, since for the second inequality to be satisfied, one needs (by construction) $\mu_{30} \gg 1$.
Eq. (\ref{eq:Lxnu}) is a direct consequence of Eq. (\ref{eq:dotnudef}), i.e. assumes that the spin--up torque is dominated by accretion. Which, in this context, is inescapable.

Therefore Eq. (\ref{eq:Lxnu}) demonstrates that magnetars cannot be present in
systems with both $L_X > 10^{39}\ergs$ \underline{and} $\dot\nu \gtrsim 10^{-10}\rm s^{-2}$, i.e. it shows that {\sl neutron stars with magnetar field-strengths cannot be present in PULXs}.
Which is fully consistent with other observational facts, such as the absence of magnetars in binary systems (for a detailed discussion see \citealt{King0519, king0623}).

Since the magnetar--PULX model is directly contradicted by observations, one is necessarily left with the geometrically--beamed--emission option. 
This means that Eqs. (\ref{eq:superedlbeam}) and (\ref{eq:rm}) have to be completed by two equations: one providing $\dot M=\dot M(R)$ that gives $\dot M(R_{\rm M})$, the other defining the beaming factor $b$.

To describe the accretion flow \citet{King1702} (hereafter KLK17; see also \citealt{King1605} and \citealt{king0623}) used the \citet{Shakura73}, ``windy'' accretion--disc model, according to
which the local radiative flux is never larger than its Eddington value, the ``excess'' power being blown away in a disc--wind. This happens inside the spherization
radius
\begin{equation}
\label{eq:resphirr}
R_{\rm sph}= {15}\dot m \,R_g,
\end{equation}
where $\dot m=\dot M/\Mdoted$ (see \citealt{king0623}; \citealt{King1702} use the original, \citealt{Shakura73}, ``27/4'' factor, instead of the correct ``15''),
resulting in
\begin{equation}
\dot M(R) \simeq \dot m_0 \dot M_{\rm Edd}\frac{R}{R_{\rm sph}}.
\label{eq:mdotr}
\end{equation}
for ${R} < {R_{\rm sph}}$ ($\dot m_0$ is the mass-transfer rate in Eddington units). Following \citet{King0902} the beaming factor is taken to be
\begin{equation}
b \simeq \frac{73}{\dot m^2}.
\label{eq:b2}
\end{equation}
For the given values of the observed quantities $L_X$ and $\dot \nu$, KLK17 obtain the value of the mass-transfer rate $\dot M_0 = \dot M(R_{\rm sph})$, the beaming factor $b$ and the neutron star's magnetic moment $\mu$ for each PULX. The results are presented in Table \ref{tab:pulxs}.
\begin{table*}[ht]
\begin{center}
\caption{Observed and derived \citep{King1702} properties of PULXs.\\ 
{\footnotesize [Based on \citet{King0520}  (corrected in  \citealt{king0623})]}}
%\vskip2pt
{
\setlength{\tabcolsep}{1pt}
\label{tab:pulxs}
{\small
%\hfill{}
\begin{tabular}{ |l||c|c||c|c|c|c|c|c|c||} 
 \hline\hline
 Name & $L_X^{\rm max}$  [erg/s] & $\dot \nu$ [s$^{-2}$]  & $\dot m_0$  & $ b$ \ \ &  ${\bm B}$\, [G]$^1$& ${\bm R_{\rm sph}}$ [cm] &  ${\bm R_M}$ [cm] \\
 \hline\hline
 M82 ULX2 & $2.0 \times 10^{40}$  & $10^{-10}$ & 46  & 0.03 & $4.0\times 10^{10}$ & $1.1\times 10^8$ & $1.0\times 10^7$  \\
 \hline
 NGC 7793 P13 & $5.0\times 10^{39}$ & $2 \times 10^{-10}$ &  25 &0.12  &  $1.1\times 10^{11}$ & $5.8\times 10^7$ & $1.6\times 10^7$  \\
 \hline
 NGC5907 ULX1 &  $\sim 10^{41}$ & $3.8 \times 10^{-9}$ \ \ & 95 &0.01 & $9.4\times 10^{12}$ & $2.2\times 10^8$ & $1.1\times 10^8 $  \\
 \hline
 NGC300 ULX1 & $4.7\times 10^{39}$ &  $5.6 \times 10^{-10}$ & 24 & 0.13  & $5.3\times 10^{11}$ & $5.5\times 10^7$ & $3.2 \times 10^7$ \\
 \hline
 M51 ULX7 & $7\times 10^{39}$ & $2.8 \times 10^{-10}$& 28 & 0.09 & $1.9\times 10^{11}$  &  $6.4\times 10^7$ & $2.0\times 10^7$   \\
 \hline
 NGC 1313 X-2 & $2 \times 10^{40}$ & $ 1.2 \times 10^{-10}$ & 46  & 0.03 & $5.3 \times 10^{10}$ & $1.1 \times 10^8$ & $1.8 \times 10^6$  \\
 \hline\hline
 SMC X-3 & $ 2.5 \times 10^{39}$ & $6.9 \times 10^{-11}$& 18 &0.23 & $2.3\times 10^{10}$ & $4.1 \times 10^7$ &  $7.1 \times 10^6$ \\
 \hline
 NGC 2403 ULX & $1.2 \times 10^{39} $ &   $3.4 \times 10^{-10}$& 13 & 0.43 & $2.5\times 10^{11}$ & $3.0\times 10^7$ & $2.3 \times 10^7$ \\
 \hline
 Swift J0243.6+6124 & $ \gtrsim 1.5 \times 10^{39}$ (?)$^2$ &  $2.2. \times 10^{-10}$& 14  &0.37 & $1.3\times 10^{11}$  &  $3.2\times 10^7$ & $1.7\times 10^7$ \\ 
 \hline
 RXJ0209.6-7427 & $1 - 2 \times 10^{39}$ & $1.165 \times 10^{-10}$& 17 & 0.25 & $5.3 \times 10^{10}$ & $3.2 \times 10^{7}$ & $1.8\times 10^6$ \\
 \hline\hline
\end{tabular}}
}
\end{center}
%\hfill{}
%\vskip 0.2truecm 
%\caption{Model parameters of ULX pulsars}
%
\begin{itemize}
\item[]
{\footnotesize
%- Based on \citet{King0520}  (corrected in  \citealt{king0623}).\\
- Systems in the last four rows are transient and have confirmed or suspected Be--star companions.\\
$^1$- $B$ obtained from $\mu$ assuming $R_{\rm NS}=10^6\rm cm$, $q=1$ and $M_{\rm NS}=1\Msun$.\\
$^2$- System in the Galaxy: distance uncertain,}
\end{itemize}
\end{table*}
Magnetic fields have normal pulsar values $10^{10}\rm G \lesssim B \lesssim 10^{13}\rm G$ and their beaming factors are moderate.
The exception is ULX1 in NGC5907, rather strongly beamed and magnetized (but still not a magnetar). In a recent paper \citep{Furst0423} reported observations of ULX1 in NGC5907 in a low state, during a phase of spindown, and, assuming that it is in a propeller regime, deduced a magnetic field $B \approx 2.5 \times 10^{13}$ G. This value is fully consistent with the result of the KLK17 model: $9.4\times 10^{12}$ G (Table \ref{tab:pulxs}), which was obtained assuming $q=1$ and $m=1$. Using $q=0.5$ and $m=1.4$, say, the resulting field is $ 1.6\times 10^{13}$ G. One should stress, however, that for such a field, neither the KLK17 nor the \citet{Furst0423} calculations are self-consistent because they do not take into account the fact that for such a high field the Eddington luminosity is no longer critical (Eq. \ref{eq:Lcrit}). But even then, ULX1 in NGC 5907 is still super-critical. Iterating, one can put the field $1.0\times 10^{13}$ G back into the KLK17 equations and obtain a solution which takes into account the fact that the critical luminosity is 43 times larger (Eq. \ref{eq:Lcrit}) than the Eddington luminosity. Then for ULX1 in NGC 5970 $L/L_{\rm crit} = 16.6$. One then obtains a solution with $\dot m_0 = 730$ ($\sim 7$ times larger than with $L_{\rm crit}={\Ledd}$), and a beaming factor $b=0.25$. 

It has been claimed often that beaming suppresses pulsations, however, \citet{King0520} pointed out that the 
magnetic axis of a neutron-star accretor is not necessarily aligned with the disc (i.e. funnel) axis, and that it is very common for the neutron star spin to be misaligned from the binary orbit defining the accretion disc plane (see \citealt{Rankin1123} for a recent example).
When these three axes are not aligned the system appears as a PULX since when the neutron-star spin axis is strongly misaligned from
the central disc axis at the spherization radius, large polar caps produce the sinusoidal pulse light curves observed in
pulsing ULXs because a significant part of the pulsed emission can escape without scattering, giving a large pulse fraction (see \citealt{King0520} for details). 

Since neutron stars in X--ray binaries have magnetic fields  spanning the range from  $10^8$ G to
several $10^{13}$ G \citep{Revnivtsev18}, PULXs are normal XRPs at a special phase of the evolution of their parent binary systems, as suggested a long time ago by \citet{King0501}.
\begin{figure}[ht]
\begin{center}
\includegraphics[width=8.0cm,height=6.0cm,angle=0]{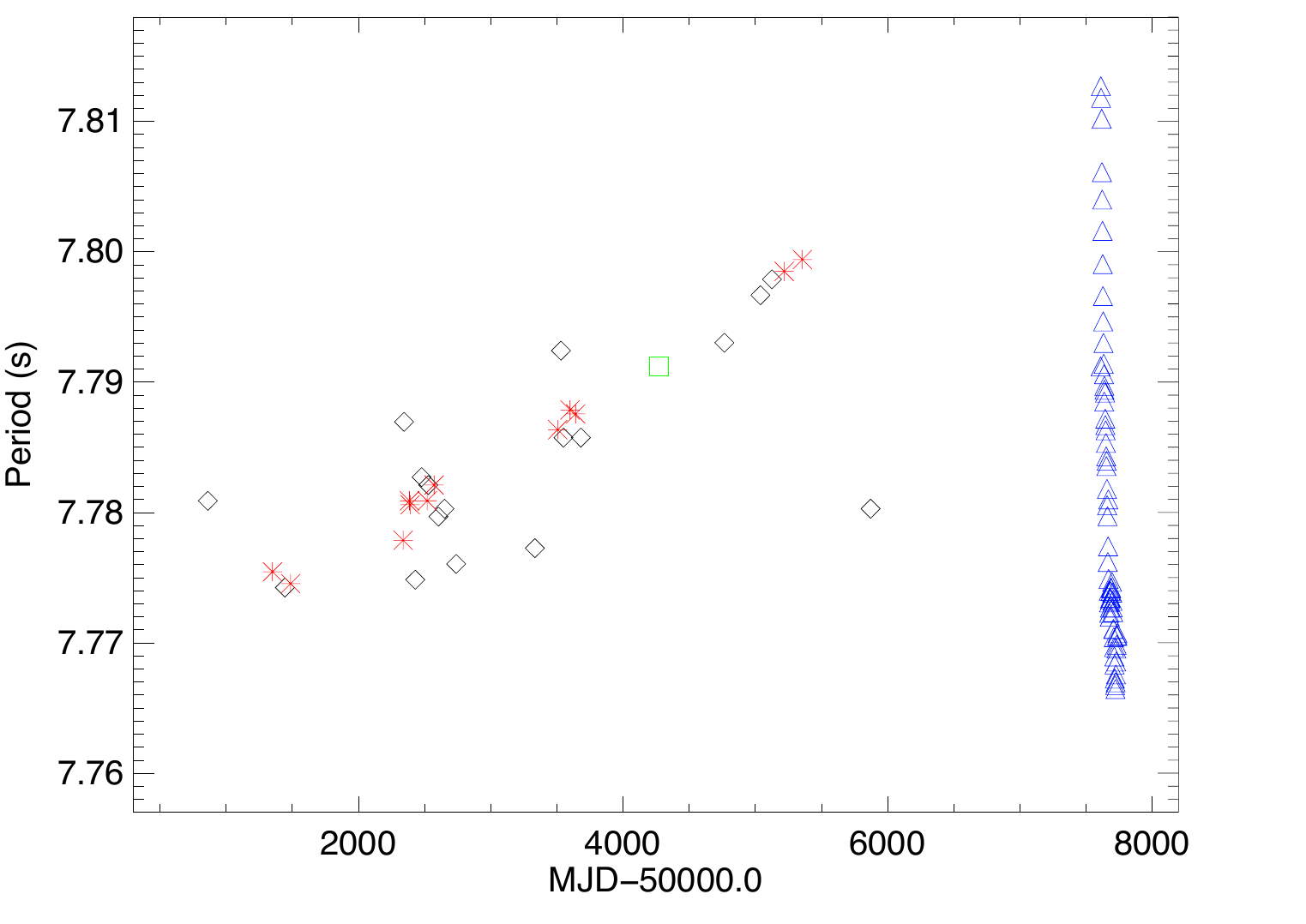}
\caption{X-ray derived pulsed period history of SMC X-3. Black diamonds
and red stars denote RXTE period detections above the 99 and 99.99 per cent
confidence levels respectively. Blue triangles denote Swift detections of the
pulse period during the current outburst. A single XMM Newton detection
at MJD 54274 was found in the literature and is denoted by a green square. ({\sl From \citealt{Townsend1117}}.)}
\label{fig:smcx3}
\end{center}
\end{figure}

This conclusion is beautifully confirmed by the Be-X binaries that become PULXs only at a certain phase of their orbital evolution. 
The binary SMC X-3 also illustrates how, during the ULX phase, the neutron--star spin evolution becomes dominated by
the accretion torque, as assumed in the KLK17 model.
Fig.~\ref{fig:smcx3} \citep{Townsend1117} shows the spin-down observed in this system, right up to the beginning of a giant outburst on MJD 57599, when significant spin-up is observed. \citet{Townsend1117} deduce from the SMC X-3 spin history that the angular momentum transferred by material accreted during the 5--month giant--outburst was larger than the angular momentum lost by magnetic braking over the previous 18 years. The long-term spin-down rate of SMC X-3 is roughly 500 times lower than the spin-{up} rate  seen during the giant outburst, showing the significantly larger torques present during this outburst. Even during previous Type I outbursts recorded by \textit{RXTE}, the spin period seemed to continue increasing under low levels of accretion. But during the giant outburst, the spin-up rate is tightly correlated with the X-ray luminosity during the super-Eddington phase \citep{Weng0717}. In other words, in PULXs the spin-up rate is strongly correlated with the X-ray luminosity both in time and over the population.

\citet{Weng0717} and \citet{Tsygankov0917} deduce the value of the magnetic field in SMC~X-3. The first team gets $6.8 \times 10^{12} \rm G$, the second $\sim 1 \times 10^{12} \rm G$, both values well below the magnetar--strength. The KLK17 model gives a lower value of $2.3\times 10^{10}\rm G$. But \citet{Tsygankov0917} use the \citet{Ghosh1179} model describing the accretion-disc -- magnetosphere interaction. Although widely used, this model is known to use very unrealistic assumptions as mentioned in Sect. \ref{sec:MGI}. On the other hand \citet{Weng0717} use a ``simple'' model describing the presumed spin--equilibrium of the system. \citet{Townsend1117} use the \citet{Ghosh1179} framework for the description the orbital spin evolution and do not get satisfactory results. In view of this, the discrepancies between these various methods of magnetic--field determination do not seem to be a serious problem. Neither of them lead to the conclusion that SMC X-3 contains a magnetar. \citet{Tsygankov0917} claim that close to the neutron-star surface the magnetic field may contain stronger components than those of the dipole. But, of course, what counts at the magnetosphere is the dipole.

Quite recently, \citet{Veledina0323} found through X-ray polarimetry observations that the WR X-ray binary Cyg X-3 is a ULX with a beaming factor\footnote{I am using here the symbol ``b'' as defined in this chapter; \citealt{Veledina0323} use $b$ to denote our $1/b$.}$b=0.02$, which corresponds to an Eddington factor $\dot m = 69$ (\citealt{Lasota1223}), but seen from the side. This system is supposed to contain a black hole. Puzzlingly, the authors of this paper do not mention the famous source SS433, which until now had been the best documented case of a ``sideways--seen'' ULX, nor do they cite any paper on the beamed--radiation interpretation of the ULXs (see the epigraph of this chapter). 

\subsection{Transient ultraluminous X-ray sources}
\label{sec:TULX}

It has  recently been established that many, and most probably most, ULXs are transient (\citealt{Brightman0523}). As mentioned above, some of the ULX transients are Be--X transient systems that occasionally become super--Eddington. Although the peak luminosities of the transient sources in \citet{Brightman0523}: $L \sim 2 -4 \times 10^{39}\ergs$ are similar to those of Be-ULXs ($\leq 4 \times 10^{39}\ergs $; see, \citealt{king0623}), their rise to outburst peak is faster than that observed in Be-ULX sources and three regions of their occurrence show varying ages of the possible stellar counterparts, while Be stars are massive and young. The transient ULX in the galaxy M51 reached a luminosity of $\sim 10^{40} \ergs$ \citep{Brightman0620}, which rather excludes a Be-ULX source. It is therefore worth trying to apply the DIM to transient ULXs, especially since it has already been successfully applied to the description of the lightcurve of XT1 in M51 \citep{Hameury1120}.
\begin{figure}[ht]
\begin{center}
\includegraphics[width=10.0cm,height=6cm,angle=0]{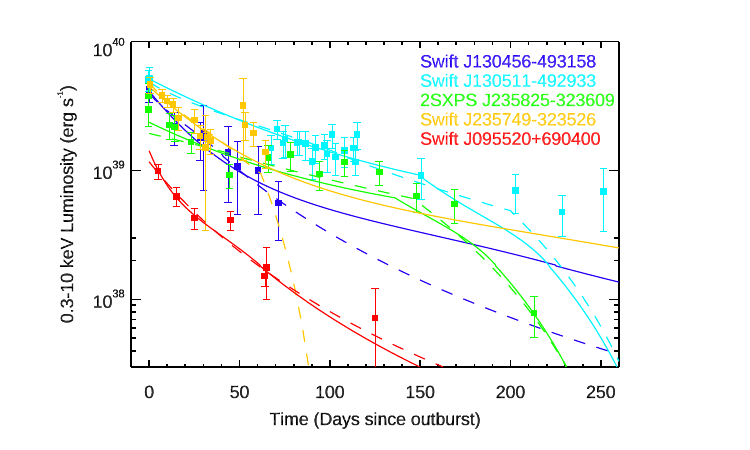}
\caption{Lightcurves of 5 transients ULXs
fitted with the disk instability model of \citet{Hameury1120}. Upper limits are omitted in the plot for
clarity. The solid lines represent the model assuming a $1.4\Msun$ accretor, whereas the dashed lines represent a system with a compact accreting body with $10\Msun$. ({\sl From \citealt{Brightman0523}.})}
\label{fig:ulxtrans}
\end{center}
\end{figure}
\citet{Brightman0523} used the formul{\ae} from Sec. \ref{sec:XRT} to fit the lightcurves of the five transient ULXs described in this paper. The results are shown in Fig. \ref{fig:ulxtrans}.

Neither the accretor masses nor the orbital parameters of these transient ULXs are known, which precludes a univocal determination of $\alpha_h$. Also,  one can determine both $\Delta t_1$ and $ t_0^{\prime}$ only in two cases, because in the other three cases a change of slope has not been observed, which leaves us with only an upper limit on the hot--disc viscosity parameter. The fits correspond to rather high values of $\alpha$: $0.3 \lesssim \alpha_h \lesssim 7$ (Brightman et al. 2023). The highest $\alpha$--value determined is 1.39, for a $10\Msun$ accretor, but the fit for the same system with $1.4\Msun$ gives 0.37. High values ($> 0.2$) were also determined by \citet{Tetarenko0218} for sub--Eddington transient outbursts of black-hole X-ray binaries.

\section{AGN}
\label{sec:AGN}

Almost 40 years ago, Martin Rees ended his seminal AGN review \citep{Rees0184} with the words:
``There has been progress toward a consensus, in that some bizarre ideas that could be seriously discussed a decade ago have been generally discarded. But if we compare present ideas with the most insightful proposals advanced when quasars were first discovered 20 years ago (such proposals being selected, of course, with benefit of hindsight), progress indeed seems meager. It is especially instructive to read \citet{Zeldovich0465} paper entitled `The Mass of Quasi-Stellar Objects'. In this paper, on the basis of early data on 3C 273, they conjectured the following: (a) Radiation pressure perhaps balances gravity, so the central mass is $10^8\Msun$. (b) For a likely
efficiency of $10\%$, the accretion rate would be $3\msy$ (c) The radiation
would come from an effective ``photosphere'' at a radius  $2 \times 10^{15}$cm
(i.e. $\gg R_g$), outside of which line opacity would cause radiation  to drive a wind. (d) The accretion may be self-regulatory, with a characteristic  time
scale of 3 yrs. These suggestions  accord  with  the  ideas  that  remain
popular  today, and  we cannot  yet make many firmly based statements that are more specific.''

When in 2023 one reads the recent assessment, ``In marked contrast to models of accretion discs around stellar mass black holes, neutron stars, and in cataclysmic variables, existing theoretical models of accretion discs around supermassive black holes do a very poor job of explaining, never mind predicting, the observed properties of luminous active galactic nuclei." (\citealt{Blaes0121}), one is tempted to conclude that during the last 60 years, theoretical progress in the field was still meagre (see also an older, but still valid diagnosis in \citealt{Antonucci0313}). 
\begin{figure}[ht]
    \centering
    \includegraphics[width=0.45\textwidth]{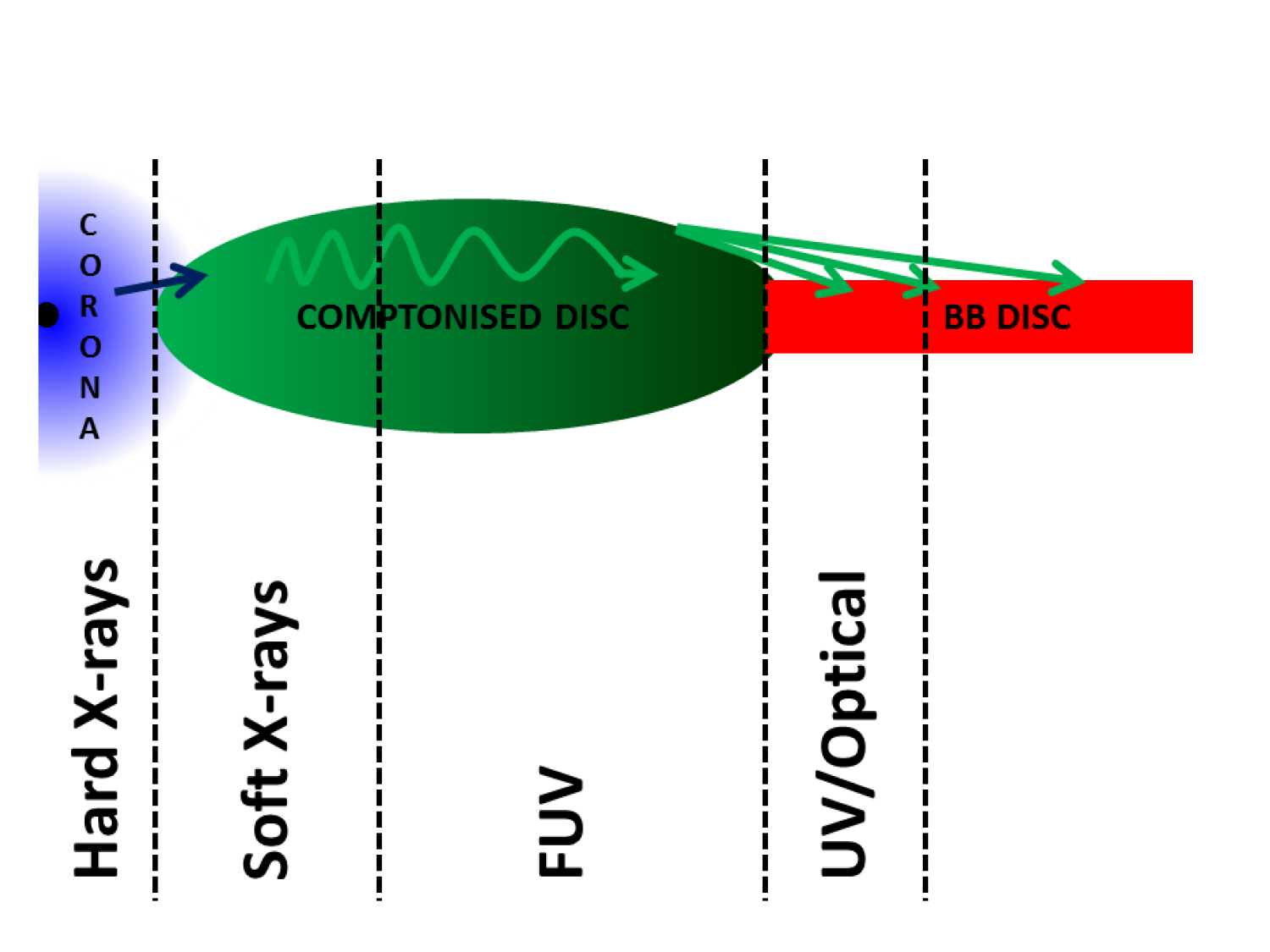}
    \includegraphics[width=0.45\textwidth]{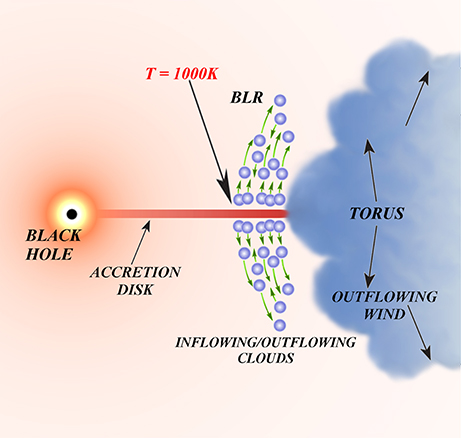}
    \caption{Left: Structure of an AGN inner accretion flow according to \citet{Gardner0917}; Right: the full AGN flow according to \citet{Czerny0617}.}
    \label{fig:agnstr}
\end{figure}
One of the reasons for this lack of progress could be deduced from the two panels on Figure \ref{fig:agnstr}.
These types of figures might be thought to illustrate what are often called ``toy models'', even if these two particular
examples are not supposed to belong to this category. Nevertheless they correspond to what could be called ``Lego models'',
since they join together various, more or less physical, models as if they were Lego bricks taken from different sets.
Is it the fault of astrophysicists, as suggested by \citet{Antonucci0313}, or are AGNs too complex and too distant
to be understood with the theoretical and observational tools available to us? Probably both.

As an illustration, I will address two problems with applying the standard disc model to AGNs: the presumed disc size and the variability
timescales.

\subsubsection{Disc radius}
The disc size can be defined as corresponding to the radius $R_\lambda$ at which the disc temperature matches the wavelength $\lambda$:
\begin{equation}
kT(R_\lambda) = hc/\lambda,
\end{equation}
i.e., 
\begin{equation}
\label{eq:rlambda}
R_{\lambda}=\left[\frac{45 G \lambda^{4} M \dot{M}}{16 \pi^{6} h c^{2}}\right]^{1 / 3}=2.1 \times 10^{15}\left(\frac{\lambda}{\mu \mathrm{m}}\right)^{4 / 3}m_8^{2 / 3}\left(\frac{L}{\eta L_{E}}\right)^{1 / 3} \mathrm{cm}.
\end{equation}
$\lambda$ is the wavelength in the rest-frame of the AGN and $M=m_8 10^8\Msun$. 
Thus the prediction of the thin-disc model is that (for a given Eddington ratio) the size of the disc satisfies the relation $R \sim M^{2/3}$, which is confirmed by observations \citep{Morgan0410} and that $R\sim \lambda^{4/3}$.
Determinations of disc  sizes through microlensing and reverberation mapping give values several times larger than those expected from Eq. (\ref{eq:rlambda}).
For example, for the AGN MCG~08-11-011 \citet{Fian0423} obtain  lags that are larger by a factor of $\sim 3 - 7$ than predictions based on the standard thin--disc model .
They also detect a size-wavelength relation significantly steeper than predicted by the model: $R\sim \lambda^{4.74}$.
However, the derivation of Eq. (\ref{eq:rlambda}) assumes that emission at wavelength $\lambda$ originates solely at radius $R_\lambda$, while in real discs
this emission also comes from other radii. Therefore a more appropriate size for comparison
with observations would be a flux--weighted mean radius $R_\lambda= {\mathcal{X}}R_\lambda$ with $ {\mathcal{X}} \approx 2 - 3$. Also, the formula for the radius assumes its stationarity, but if the disc
variability is taken into account this factor could even be $\sim 5$. In addition, $R_\lambda$ depends on the black hole mass and the accretion rate (through radiative efficiency), so uncertainties in the values of these quantities might influence the comparison of the model-size with observations. 

However, the main weak point of Eq. (\ref{eq:rlambda}) is the assumption that every ring of the putative disc radiates like a black body, and what is observed
is the effective temperature. This is not the case with discs in cataclysmic variables (see Sect. \ref{sec:CV}), and it is not the case here. The emitted spectrum depends on the
details of the disc's vertical structure, which are not well known, and what is observed is the colour temperature
\begin{equation}
    T(r) = f_{\rm col}\left( \Teff (r)\right)\Teff(r),
\end{equation}
where $f_{\rm col}\left( \Teff (r)\right) \geq 1$ is the colour temperature correction. 
This formula was used by \citet{Zdziarski1122} to reevaluate AGN disc radii.
Instead of the disc radius, they calculate the half-light radius, $R_{1/2,\nu}$, at which half
of the emission at the frequency $\nu$ is emitted inside $R_{1/2,\nu}$.
They apply the disc model to the quasar SDSS 0924+0219, whose half-light radius, at $\nu \approx 4.8 \times 10^{14}$ Hz, from
microlensing, is $R_{1/2,\nu} \approx 150 R_g$ \citep{Morgan0410}. Using two different colour corrections, \citet{Zdziarski1122}
obtain the same result: $R_{1/2,\nu} \approx 107R_g$. Still too short, but just by a factor of 1.4. When the effect of the disc's inner truncation
and disc winds is added, the half--light radius rises to 128$R_g$. The authors mention that AGN discs are supposed to have X-ray coronas, whose 
effect would be to multiply the disc size by a factor of $(1 - f_c)^{-1/2}$, where $f_c < 1$ is the fraction of the disc
emission that is lost due both to its  covering by a corona and to the coronal dissipation. This might be so, but by adding all these ingredients (except for the colour correction) we end up trying to force a more complex system into the shape of a thin accretion disc, just as Ptolemeus was attempting to make an ellipse into a circle with ``equants'' and ``epicycles'

Therefore, to the question: are observations of AGN disc sizes compatible with the presence there of a stationary,  geometrically thin,
optically thick Keplerian disc\footnote{This does {\sl not} have to be a \citet{Shakura73} disc which is a power-law solution assuming a specific form of the viscosity torque and opacity law. The result is much more general.} ?, the answer seems to be: ``no''. It is possible, however, that the disc size problem is a red herring because of the large intrinsic reddening of AGNs not being properly taken into account, which leads to underestimating their bolometric luminosities ($L$ and $\dot M$ in Eq. \ref{eq:rlambda}, \citealt{Gaskell0517,Gaskell0323}). Still, it is not clear that taking into account the intrinsic reddening results in a self-consistent model of AGNs accreting from stationary geometrically thin discs. (see e.g. \citealt{Antonucci0923}). In any case, the thin--disc model in AGNs has to face other serious challenges.

\subsubsection{Timescales}

The most conspicuous challenge to the thin disc model is probably the extreme variability observed in so--called  changing
look AGNs (CLAGNs). In these sources, the UV--optical continuum and broad emission line spectral components appear or disappear on timescales of months to years. In the case of Seyferts, this corresponds to a transition between AGN spectra that contain broad emission lines (i.e. Seyfert 1 type) and those with only narrow lines (Seyfert~2 type).
For thin discs in the gas--pressure dominated regime the viscous time is
\begin{equation}
t_{\rm vis} \approx 1.5 \times 10^4 \alpha^{-1}\,T^{1/2}_4\,m_8\,r^{1/2} \rm \,yr,
\end{equation}
so these variability timescales are  related rather to the thermal timescale
\begin{equation}
\label{eq:ttherm}
t_{\rm th}=\frac{1}{\alpha}t_{\rm dyn}= 1.4 \times 10^4 \alpha_{0.1}^{-1}m_8 r^{3/2} \,\rm s,
\end{equation}
but we still need to explain how huge changes in luminosities can be triggered without affecting the accretion rate.

\citet{Jiang0920} remark that there is a very important difference between accretion discs in bright
AGNs and those in cataclysmic variables and X-ray binaries: discs around supermassive black holes ``have thermal pressures that are hugely dominated by
radiation pressure'', which in the case of the latter could explain the CLAGN phenomenon.

At first glance this assertion might seem surprising, since (when $\kappa_{\rm abs} \gg \kappa_{\rm es}$), according to the \citet{Shakura73} model
\begin{equation}
\frac{P_g}{P_r}= 0.32\, \alpha^{-1/10}m_8^{-1/10}r^{3/8}\dot m^{-7/20}f^{-7/20},
\label{eq:pgpr}
\end{equation}
so that this ratio, close to a $10^8\Msun$ black hole, say, is only 6 times smaller than it would be near a neutron star. On the other hand, only hugely super--Eddington accretion rates ($\dot m \gtrsim 50$) would make $P_r \gg P_g$, 

However, true absorption dominates the opacities only for 
\begin{equation}
r_{\rm Res} > 7400 \,\dot m^{2/3},
\label{eq:rkappases}
\end{equation}
where I used the Kramers opacities
\begin{equation}
\kappa_{\rm R}= 5\times 10^{24}\rho T^{-7/2} = 5\times 10^{-4}\dot m^{-1/2} r^{3/4} \rm cm^2g^{-1},
\label{eq:kappass}
\end{equation}
so it seems that, as in the original \citet{Shakura73} solution, the $P_r \gg P_g$ regime can exist only when  $\kappa_{\rm es} \gg \kappa_{\rm abs}$.
This will happen for radii
\begin{equation}
 r \lesssim 100 \, \alpha^{2/21} \dot m^{16/21} m_8^{2/21},
\end{equation}
and there, i.e. for radii less than about few tens of $R_S$
\begin{equation}
    \dfrac{P_r}{P_g} \propto \alpha^{1/4}m^{1/4} \dot m^{2}r^{-21/8},
\end{equation}
so that,  at comparable Eddington (accretion--rate) factors, radiation pressure in discs around supermassive BHs can indeed be 100 times larger than
in discs around NS and stellar--mass BH.
\begin{figure}[ht]
    \centering
    \includegraphics{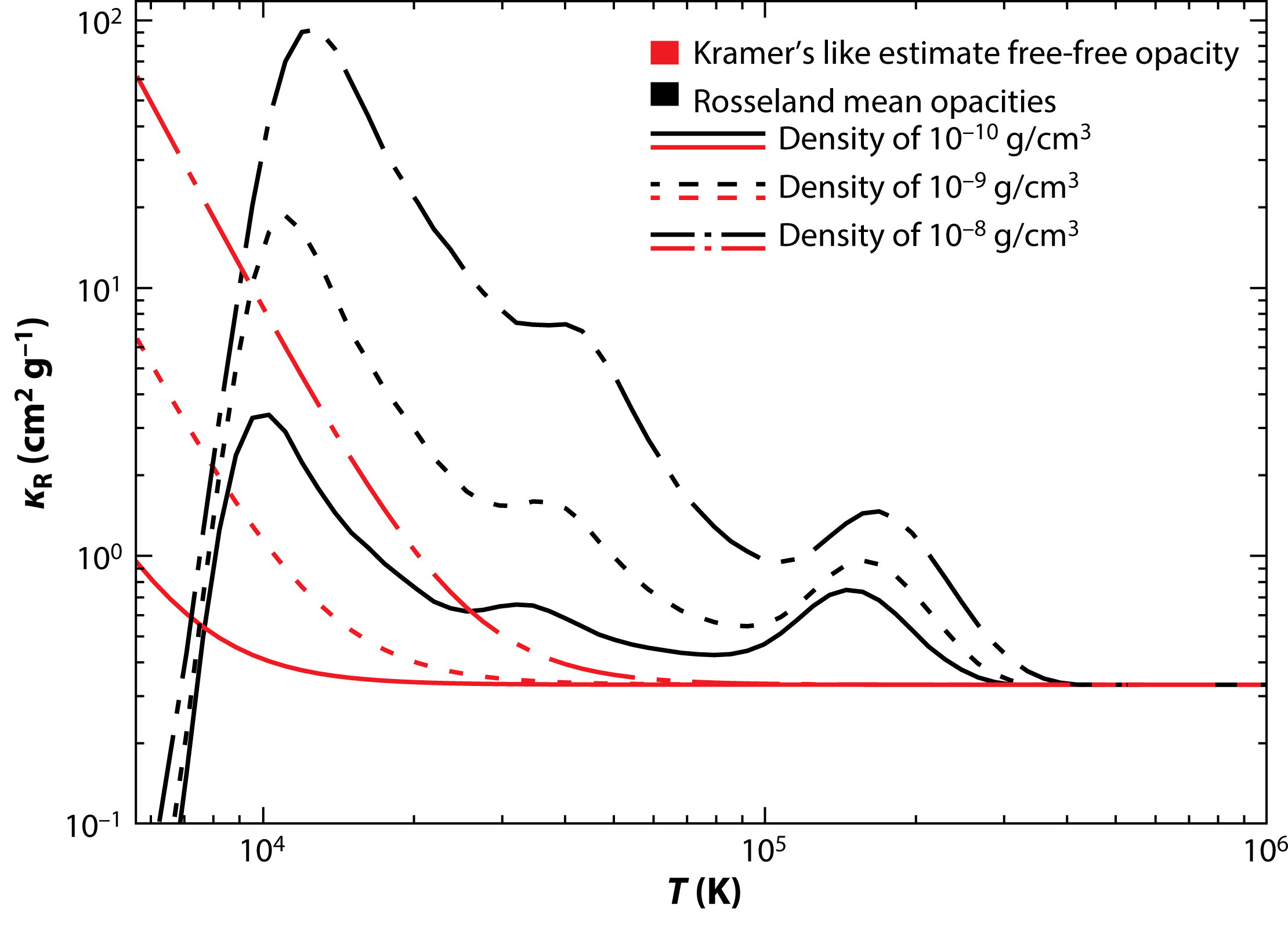}
    \caption{Comparison of the sum of electron scattering opacity and Rosseland mean opacities
from the OPAL project (black curves) with a Kramers-like estimate free-free opacity (red
curves). Lines correspond to densities of $10^{-8}\gcm$ (solid), $10^{-9}\gcm$ (dotted), and $10^{-10}\gcm$
(dash-dotted). The enhancement at $\sim 1.8\times 10^5$K corresponds to the Fe opacity bump.({\sl From \citealt{Davis0820}}).}
    \label{fig:opacities}
\end{figure}

But the real difference between accretion at high rates onto supermassive BH and stellar--mass compact bodies is the temperature of the inner parts of the accreting matter. In the case of NS and stellar--mass BHs the (effective) temperature is $\gtrsim 10^8$K, for white dwarfs it is $\gtrsim 10^4$K, while for supermassive black holes it is  $\gtrsim 10^5$K, which corresponds to UV radiation. In addition, the density is much lower than in discs around stellar--mass accretors: $\rho < 10^{-8}\gcm$. In such conditions the inner region of the accretion disc can be in a regime {\sl non--existing} in the standard,
three--zone Shakura--Sunyaev model: $\kappa_{\rm abs} > \kappa_{\rm es}$ and $P_r \gg P_{g}$. The reason is illustrated in Fig. \ref{fig:opacities}, taken from \citet{Davis0820}.
Therefore, for typical conditions in the inner--disc region of bright AGNs, the Rosseland mean opacity is expected to be larger than the electron scattering
value. \citet{Jiang0920} show that the iron opacity bump (around $1.8 \times 10^5$K) causes the disc to be convectively unstable. Their simulations show that turbulence generated by convection increases the disc thickness due to additional turbulent--pressure support and enhances the local angular momentum transport. They find that this also results in strong fluctuations in surface density and heating of the disc. When the  opacity drops with increasing temperature, the convection is suppressed, the disc cools down and the whole cycle
repeats again. As a result, the disc scale height strongly oscillates, causing luminosity variations of more than a factor of $\approx 3 - 6$ on a  timescale of a few years. The authors propose that this is the physical mechanism which explains AGN variability with a wide range of amplitudes over a timescale of years to decades.

This is, however, doubtful, as demonstrated by the example of  Mrk 590, a nearby CLAGN \citep{Lawther0323}. In its high state, it is a Seyfert 1, i.e. a moderate--accretion rate source, but it shows X-ray and UV variability amplitudes higher than those typically observed in steady-state AGNs of this type. Its variability is similar to that of highly accreting AGNs, i.e. quasars. The characteristic timescale of the Mrk 590 flares is $\sim 100$ d. Since its mass is $4.75 \times 10^7 \Msun$, this corresponds to the thermal timescale of its putative accretion discs (see Eq. \ref{eq:ttherm} and Fig. 10 in \citealt{Lawther0323}). However, with an accretion rate $\dot m = 0.05$ at maximum, it is unlikely that the \citet{Jiang0920} mechanism is here at work . At such accretion rates the disc is radiation--pressure dominated only at its innermost tip. The X-ray and UV flares in Mrk 590 have a complex spacetime structure, which seems to exclude a simple disc--like structure. Incidentally, the X-rays in this source irradiate the UV--emitting region, which might suppress convection \citep{Dubus9902}, if any.

Recently, \citet{Yang0323} suggested that changes observed in changing-look quasars (CLQ) are due
to changing accretion rates, with the multiwavelength emission varying accordingly. In this they find ``promising analogies to the accretion states of X-ray binaries''.
One should, however, be very careful with the interpretation of apparent analogies in the behaviour of accreting systems at different mass and space scales. For example,
it has been shown that the hystereses observed in the hardness--intensity diagrams of LMXBs (in X--rays) and of dwarf--nov{{\ae}} (in optical vs EUV/X-ray), despite their apparent similarity, are due to two completely different mechanisms (\citealt{Hameury0417}; see also \citealt{Hameury0420}). In the case of apparent analogies between CLQs and binary X--ray transients, one should stress that in the latter, the viscous timescale in the relevant disc region is much shorter than the observed variability times, in striking contrast to CLQs and CLAGNs, so that the existence of a common, or similar, mechanism explaining both these classes of phenomena is rather doubtful.

Interestingly, at $\dot m \lesssim 0.05$, the accretion disc of Mrk 590 should be subject to the dwarf--nova--type thermal--viscous instability \citep{Hameury0609,Lasota22}, but the observed variability of this AGN is nothing like the lightcurves produced by this mechanism (see Fig. 7 in \citealt{Lasota22}). The reason is that, although heating and cooling fronts propagate throughout the disc, their movement is too fast to  significantly affect the disc's density.
The front propagation timescale is
 \begin{equation}
t_{\rm front} \approx \frac{R}{\alpha c_{\rm s}} = \frac{R}{H} t_{\rm th},
 \end{equation}
where $t_{\rm th}$ is the thermal timescale. Hence $t_{\rm front}$ is 
shorter than the viscous time $t_{\rm visc} = (R/H)^2 t_{\rm th}$ by
a factor of $H/R$, i.e. by several orders of magnitude, since (in a gas--pressure dominated disc)
\begin{equation}
    \frac{H}{R} \approx \frac{c_s}{v_K}= 5.5 \times 10^{-5} T_4^{1/2} r^{1/2},
\end{equation}
where $c_s$ and $v_K$ is the Keplerian speed. Another difference between the discs in binaries and in AGNs is that in the former, the optical emission region is at $r \gtrsim 1000$, while in the latter it is at $\gtrsim 10$, much deeper in the gravity potential well, if the AGNs have stationary, flat, geometrically thin, optically thick Keplerian accretion discs. Which is less than certain. But there are alternatives: see e.g. chapter 5 in \citet{KingSMBH}.

\section*{Acknowledgements}
I am grateful to Robert Antonucci, Mitchell Begelman, Jean-Marie Hameury and Andrew King for inspiring comments, criticism  and discussions. I thank Martin Gaskell, Jonathan Katz, Galina Lipunova and Andrzej Zdziarski for their valuable advice and comments on the first posted version of this article. I also thank an anonymous referee for the very detailed report which helped me to correct several errors in the original manuscript.
I acknowledge with thanks the variable star observations from the AAVSO International Database contributed by observers worldwide and used in this research.
%\citet{Lasota22}

\bigskip
\bigskip

\bibliographystyle{apalike}
\linespread{0.25}
\renewcommand\bibname{References}
\footnotesize
\bibliography{Lasota-100thyear}

\end{document}